\pdfoutput=1

\documentclass[11pt]{article}

\usepackage{ACL2024}

\usepackage{times}
\usepackage{latexsym}
\usepackage{arydshln}
\usepackage{graphicx}
\usepackage{subcaption}
\usepackage{booktabs,arydshln}
\usepackage{amsmath}
\usepackage{enumitem}
\usepackage{multirow}
\usepackage{cleveref}
\usepackage[export]{adjustbox}
\usepackage[utf8]{inputenc}
\usepackage{xurl}
\usepackage{tabularray}
\usepackage{siunitx}
\usepackage{amssymb}
\usepackage{pifont}

\newcommand{\circa}{{\raise.17ex\hbox{$\scriptstyle\sim$}}}

\usepackage{array}
\newcolumntype{L}[1]{>{\raggedright\let\newline\\\arraybackslash\hspace{0pt}}m{#1}}
\newcolumntype{C}[1]{>{\centering\let\newline\\\arraybackslash\hspace{0pt}}m{#1}}
\newcolumntype{R}[1]{>{\raggedleft\let\newline\\\arraybackslash\hspace{0pt}}m{#1}}

\usepackage[T1]{fontenc}

\usepackage[utf8]{inputenc}

\usepackage{microtype}

\usepackage[colorinlistoftodos]{todonotes}
\usepackage{cleveref}
\usepackage{listings}

\makeatletter
\newcommand*\iftodonotes{\if@todonotes@disabled\expandafter\@secondoftwo\else\expandafter\@firstoftwo\fi} 
\makeatother

\crefname{section}{\S}{\S\S} 
\Crefname{section}{\S}{\S\S} 
\crefname{table}{Tab.}{Tables}
\crefname{figure}{Fig.}{Figures}
\crefname{algorithm}{Algorithm}{}
\crefname{equation}{eq.}{}
\crefname{appendix}{App.}{}
\crefname{lstlisting}{listing}{listings}
\Crefname{lstlisting}{Listing}{Listings}
\crefformat{section}{\S#2#1#3} 

\definecolor{KUPetrol}{RGB}{0,120,148} 

\definecolor{KUBlue}{RGB}{33,92,175} 
\definecolor{KUGreen}{RGB}{98,115,19} 
\definecolor{KUPurpleDark}{RGB}{140,10,89} 
\definecolor{KUPurple}{RGB}{163,7,116} 
\definecolor{KUGray}{RGB}{111,111,111} 
\definecolor{KURed}{RGB}{183,53,45} 
\definecolor{KUPetrol}{RGB}{0,120,148} 
\definecolor{KUBronze}{RGB}{142,103,19} 

\newtoggle{color-macro}
\settoggle{color-macro}{true} 

\iftoggle{color-macro}{\colorlet{MacroColor}{KUPetrol}
}{
\colorlet{MacroColor}{black}
}

\iftoggle{color-macro}{
\colorlet{TokenColor}{KUBronze}
}{
\colorlet{TokenColor}{black}
}

\iftoggle{color-macro}{
\colorlet{MathSubColor}{KUPurple}
}{
\colorlet{MathSubColor}{black}
}

\iftoggle{color-macro}{
\colorlet{RedditColor}{black}
}{
\colorlet{RedditColor}{black}
}

\iftoggle{color-macro}{
\colorlet{SchwartzProbColor}{KUBlue}
}{
\colorlet{SchwartzProbColor}{black}
}

\iftoggle{color-macro}{
\colorlet{IColor}{KUGray}
}{
\colorlet{IColor}{black}
}

\newcommand{\mymacro}[1]{{\color{MacroColor} #1}}
\newcommand{\TokenMacro}[1]{{\color{TokenColor} #1}}
\newcommand{\RedditMacro}[1]{{\color{RedditColor} #1}}
\newcommand{\MathSubMacro}[1]{{\color{MathSubColor} #1}}
\newcommand{\SchwartzProbMacro}[1]{{\color{SchwartzProbColor} #1}}

\newcommand{\specialtoken}[1]{\TokenMacro{\texttt{#1}}}
\newcommand{\reddit}{\RedditMacro{Reddit}}
\newcommand{\mathsubreddit}{\MathSubMacro{\mathcal{S}}}
\newcommand{\schwartzvec}{\SchwartzProbMacro{\vu}}
\newcommand{\similarityFunction}[1]{\mymacro{\sigma}_\TokenMacro{\text{#1}}}

\usepackage[T1]{fontenc}

\newcommand{\subreddit}[1]{\href{https://www.reddit.com/r/#1}{\fontfamily{qcr}\selectfont{r/#1}}}

\newcommand{\subredditpositive}[1]{\href{https://www.reddit.com/r/#1}{\textcolor{OliveGreen}{\fontfamily{qcr}\selectfont{r/#1}}}}

\newcommand{\subredditnegative}[1]{\href{https://www.reddit.com/r/#1}{\textcolor{BrickRed}{\fontfamily{qcr}\selectfont{r/#1}}}}

\newcommand{\schwartzvalue}[1]{\textit{#1}}

\newcommand{\mostSimilarSubreddit}[1]{\overline{\mathsubreddit{}}_{\text{#1}}}

\usepackage{amsmath,amsfonts,bm}

\newcommand{\newterm}[1]{{\bf #1}}

\def\eqref#1{equation~\ref{#1}}

\def\1{\bm{1}}

\def\ve{{\bm{e}}}

\def\vu{{\bm{u}}}

\def\evi{{i}}

\def\evk{{k}}

\DeclareMathAlphabet{\mathsfit}{\encodingdefault}{\sfdefault}{m}{sl}
\SetMathAlphabet{\mathsfit}{bold}{\encodingdefault}{\sfdefault}{bx}{n}

\definecolor{codegreen}{rgb}{0,0.6,0}
\definecolor{codegray}{rgb}{0.5,0.5,0.5}
\definecolor{codepurple}{rgb}{0.58,0,0.82}
\definecolor{backcolour}{rgb}{0.97,0.97,0.95}

\lstdefinestyle{mystyle}{
    backgroundcolor=\color{backcolour},   
    commentstyle=\color{codegreen},
    keywordstyle=\color{magenta},
    numberstyle=\tiny\color{codegray},
    stringstyle=\color{codepurple},
    basicstyle=\ttfamily\footnotesize,
    breakatwhitespace=true,         
    breaklines=true,                 
    captionpos=b,                    
    keepspaces=true,                 
    numbers=left,                    
    numbersep=5pt,                  
    showspaces=false,                
    showstringspaces=false,
    showtabs=false,                  
    tabsize=2
}

\title{Investigating Human Values in Online Communities}

\author{
Nadav Borenstein$^1$ \quad
Arnav Arora$^1$ \quad
Lucie-Aim\'{e}e Kaffee$^{2}$ \quad
\textbf{Isabelle Augenstein}$^{1}$ \\
$^1$University of Copenhagen \quad
$^2$Hasso Plattner Institut \quad \\
{\tt \href{mailto:nb@di.ku.dk}{nb@di.ku.dk}} \quad
{\tt \href{mailto:aar@di.ku.dk}{aar@di.ku.dk}} \quad \\
{\tt \href{mailto:lucie-aimee.kaffee@hpi.de}{lucie-aimee.kaffee@hpi.de}}  \quad 
{\tt \href{mailto:augenstein@di.ku.dk}{augenstein@di.ku.dk}} \quad 
}

\begin{document}
\maketitle


\begin{abstract}
Studying human values is instrumental for cross-cultural research, enabling a better understanding of preferences and behaviour of society at large and communities therein. To study the dynamics of communities online, we propose a method to computationally analyse values present on Reddit. Our method allows analysis at scale, complementing survey based approaches. We train a value relevance and a value polarity classifier, which we thoroughly evaluate using in-domain and out-of-domain human annotations. Using these, we automatically annotate over nine million posts across 12k subreddits with Schwartz values. Our analysis unveils both previously recorded and novel insights into the values prevalent within various online communities. For instance, we discover a very negative stance towards conformity in the Vegan and AbolishTheMonarchy subreddits. Additionally, our study of geographically specific subreddits highlights the correlation between traditional values and conservative U.S. states. Through our work, we demonstrate how our dataset and method can be used as a complementary tool for qualitative study of online communication.
\end{abstract}
\everypar{\looseness=-1}

\section{Introduction}
\label{sec:introduction}
\begin{table}[ht]
    \centering
    \fontsize{8}{9}\selectfont
\begin{tabular}{p{0.3cm}p{6.5cm}}
\toprule
Val & Subreddits \\
\midrule
AC    &   \subredditpositive{startups}, \subredditpositive{resumes}, \subredditpositive{xboxachievements}  \\
BE    &             \subredditpositive{Adoption}, \subredditnegative{BPDlovedones}, \subredditpositive{Petloss}  \\
CO     &                    \subredditnegative{policebrutality}, \subreddit{HOA}, \subredditnegative{BadNeighbors}, \\ 
HE       &  \subredditpositive{FreeCompliments}, \subredditpositive{transpositive}, \subredditpositive{cozy} \\
PO          &                                  \subredditpositive{debtfree},  \subredditpositive{geopolitics}, \subredditpositive{dividends}\\
SE       &   \subreddit{GunsAreCool}, \subredditpositive{worldevents}, \subredditpositive{CombatFootage} \\
SD &                  \subreddit{antidepressants} \subredditpositive{DebateReligion}, \subreddit{TrueUnpopularOpinion}, 
\subredditpositive{nutrition} \\
ST    & \subredditpositive{crossdressing}, \subredditpositive{Hobbies}, \subredditpositive{NailArt} \\
TR      & \subredditnegative{religion}, \subredditpositive{AskAPriest}, \subredditnegative{atheism} \\
UN   &  
\subredditpositive{AskFeminists}, 
\subredditpositive{IsraelPalestine}, 
\subredditpositive{climatechange} \\
\bottomrule

\end{tabular}  
    \caption{Subreddits with the highest expression of each of the ten Schwartz values. The stance of \textcolor{OliveGreen}{Green} subreddits towards the value is positive (above $0.2$), whereas \textcolor{BrickRed}{Red} indicates negative stance (below $-0.2)$. \textcolor{darkblue}{Blue} represents neutral.
    AC=achievement, BE=benevolence, CO=conformity, HE=hedonism, PO=power, SE=security, SD=self-direction, ST=stimulation, TR=tradition, UN=universalism.
}
\label{tab:strongest_signal_head}
\end{table}


Human Values have been a useful analysis lens for social sciences scholars \citep{ponizovskiy-etal-2020-pvd, Boyd_Wilson_Pennebaker_Kosinski_Stillwell_Mihalcea_2021, schwartz-1994-universal}. Applied to communities or individuals, they are used to study political affiliations, cultural integration, human disagreement, economic growth, and human development, among others~\citep{inglehart2020modernization}. Many frameworks exist for studying human values, including Rockeach values~\citep{rokeachNatureHumanValues1973}, Hofstede's cultural dimensions~\citep{hofstede1984culture}, and Moral Foundations Theory~\citep{graham-2013-mft}. However, studies using these value frameworks often struggle with sample sizes and rely on self-reported surveys to calculate the values of communities \citep{weld2023making}, leading to concerns about representation and the generalisation of the results to populations beyond the ones studied \citep{gerlach2021measuring, bavsnakova2016dimensions, doi:10.1177/0022022108318112}. 

Social media platforms provide unadulterated access to vast and diverse expressions of human thoughts and opinions in the form of posts, discussions, and comments. This rich data source is invaluable for investigating various aspects of human society~\citep{newell2016user,agrawal2022wallstreetbets,zomick-etal-2019-linguistic,turcan-mckeown-2019-dreaddit}. However, analysing large amounts of social media data qualitatively remains challenging.


\begin{table*}[ht]
    \centering
    
    \fontsize{8}{9}\selectfont
\begin{tabular}{lp{13cm}}
\toprule
Value & Description \\
\midrule

\schwartzvalue{Power}	& Social status and prestige, control or dominance over people and resources \\
\schwartzvalue{Achievement}	& Personal success through demonstrating competence according to social standards. \\
\schwartzvalue{Hedonism} &	Pleasure and sensuous gratification for oneself. \\
\schwartzvalue{Stimulation} &	Excitement, novelty, and challenge in life. \\
\schwartzvalue{Self-direction} &	Independent thought and action-choosing, creating, exploring. \\
\schwartzvalue{Universalism}	& Understanding, appreciation, tolerance, and protection for the welfare of all people and for nature. \\
\schwartzvalue{Benevolence}	& Preservation and enhancement of the welfare of people with whom one is in frequent personal contact. \\
\schwartzvalue{Tradition}	& Respect, commitment, and acceptance of the customs and ideas that traditional culture or religion provide. \\
\schwartzvalue{Conformity} &	Restraint of actions, inclinations, and impulses likely to upset or harm others and violate social expectations or norms. \\
\schwartzvalue{Security} &	Safety, harmony, and stability of society, of relationships, and of self. \\
\bottomrule

\end{tabular}
    \caption{The ten Schwartz values and their meaning; descriptions from \citet{schwartz-1994-universal}}\label{tab:value_description}
\end{table*}

To overcome this challenge, in this work, we provide a method for computationally analysing human values in language used on social media at scale. Our method and dataset supplies scholars with a tool for extracting high level values based insights from a large corpus of social media text, which can be used to identify interesting phenomena to qualitatively study when studying online behaviour and communities. Specifically, we follow Schwartz's Theory of Human Values~\citep{schwartz-1994-universal}, due to its wide adoption as a value framework in the social sciences as well as in Natural Language Processing, and apply it on \reddit{}. \reddit{} is conceptually based on named communities, or, \newterm{subreddits}, and studying the values they exhibit is particularly interesting for studying online behaviour as the discussions are already segregated by topic or sub-communities. For instance, by examining the \subreddit{teenagers} subreddit, we can gain insights into adolescent perspectives, while the \subreddit{ConservativeValues} allows us to study the impact of conservative values on worldviews.
 
To extract values from Reddit, we train supervised value extraction models to classify the \textit{presence} and \textit{polarity} of Schwartz values in text and apply them to Reddit posts. We then evaluate the models, validating their effectiveness and addressing limitations in prior work. Subsequently, we use these models to infer the expressed values of 9 million user posts and comments across 11,616 most popular subreddits on Reddit.

Through such a computational value analysis, we can analyse digital trace and behaviour data at scale, complementing social science studies by highlighting patterns as well as outliers needing further study. As an instance of this, our analysis demonstrates that people contributing to \subreddit{feminism} exhibit high values in  \schwartzvalue{self-direction}, as substantiated by previous studies. We also demonstrate the prevalence of high traditional values in conservative US states. 
In sum, our \textbf{contributions} are:
\begin{itemize}[noitemsep]
    \item A flexible method of conducting large-scale value relevance and polarity analysis to complement social science research on online communities; 
    \item Our analysis of values at a large scale confirms previously recorded phenomena and unveils novel insights, indicating that our method can complement traditional methods in understanding societal phenomena;
    \item We release our dataset of 12k subreddits with their corresponding Schwartz values for future analysis.\footnote{\url{https://github.com/copenlu/HumanValues}}
\end{itemize} 
Although our findings offer unique insights into the values of individuals from various cultures, 
the focus of our paper is the analysis of values present in \emph{online communities}. Hence the subject of our study is the communities themselves and how they operate, rather than the broader cultural backgrounds of their members, which we do not measure directly. Our methods are designed to complement, not replace, traditional survey-based approaches for studying values, by providing additional perspectives from digital trace data.




\section{Related Work and Background}
\label{sec:related_work}

\subsection{Schwartz's Values Framework}
\label{subsec:background}
Values represent a crucial aspect of human nature. According to ~\citet{Schwartz2012_overview}, \textit{"A person’s value priority or hierarchy profoundly affects his or her attitudes, beliefs, and traits, making it one
core component of personality."}
%
%
%
~\citet{schwartz-1994-universal} define values as (1) concepts or beliefs, that (2) pertain to desirable end states or
behaviors, (3) transcend specific situations, (4) guide selection or evaluation of
behavior and events, and (5) are ordered by relative importance. Based on this, they outline ten basic human values:~\schwartzvalue{Security, Conformity, Tradition, Benevolence, Universalism, Self-direction, Stimulation, Hedonism, Achievement, Power}. We provide descriptions of each of the values in \Cref{tab:value_description}.
The values were originally applied to measure differences in values across cultures~\cite{schwartz-1994-universal}.
The framework is suggested as a tool to study populations rather than individual people, hence it serves as a suitable tool for the analysis of online communities.
We leverage the fact that a person's identity and values are often reflected in the linguistic choices they make~\cite{Jaffe2009StanceSP, norton-lang-identity} to analyse values embedded in text.

\subsection{Values and Natural Language Processing}
Recently, there have been a number of studies exploring values and morals using NLP. In analysing values and language on social media, ~\citet{ponizovskiy-etal-2020-pvd} released a Schwartz value dictionary. 
~\citet{Boyd_Wilson_Pennebaker_Kosinski_Stillwell_Mihalcea_2021} show the promise of free-text survey response and Facebook data for value extraction.
There have also been studies exploring morals and norms in text. ~\citet{trager2022moral} release the Moral Foundations Reddit Corpus with 16k Reddit comments; ~\citet{roy-etal-2021-identifying} similarly study morality in political tweets. ~\citet{havaldar-etal-2024-building} study the presence of values in a geolocated Twitter corpus using a lexicon-based approach, finding a lack of correlation with survey data. The closest to our study is~\citet{van-der-meer-etal-2023-differences}, who train a value extraction model based on datasets from ~\citet{qiu2022valuenet,kiesel-etal-2022-identifying}. 
Using the model, they analyse values at an individual user level to understand disagreement in online discussions. However, a key limitations in their work is only detecting the \textit{presence} of a value, neglecting the \textit{polarity} of the discussion towards that value. 
In our work, we overcome this limitation by training an additional value \textbf{stance} model. Further, we perform our analysis at the community level to understand the values of communities rather than individuals, bringing it closer to the original framework outlined by Schwartz designed to understand cultural values. 
\looseness=-1

\subsection{Studying Online Communities}
\label{sec:social_media}

Much work has been done to explore Reddit and its user base. Reddit users' personalities have been studied \cite{gjurkovic-snajder-2018-reddit} as well as their mental health \cite{zomick-etal-2019-linguistic,turcan-mckeown-2019-dreaddit, chancellor-etal-2018-ohc}. Previous work has also studied Reddit by focusing on events affecting community dynamics \cite{newell2016user,agrawal2022wallstreetbets}. For instance, \citet{10.1145/3490499} examine the effect of content moderation on two controversial subreddits. \citet{10.1145/3342220.3343662} study the characteristics and differences between left-leaning and right-leaning political subreddits. Closer to our work, \citet{weld2023making} craft a taxonomy of community values and associate them with specific subreddits. However, their methodology, while high quality, involves collecting values through self-reporting questionnaires of limited size and suffers from selection bias, a limitation we address in this paper.
Finally, studies have also explored modelling community norms~\citep{park2024valuescopeunveilingimplicitnorms} and detecting their violations~\citep{cheriyan-etal-2021-so-norms}.\looseness=-1



\section{Method}
\label{sec:method}
This section details the collection and processing of our \reddit{} dataset, and our approach to training and evaluating the Schwartz values extractor model.

\subsection{Data Collection}

\begin{table}
    \centering
    \resizebox{0.5\textwidth}{!}{%
    \fontsize{10}{10}\selectfont
    \sisetup{table-format = 3.2, group-minimum-digits=3}
    \begin{tabular}{l|rrrrrr} \toprule
             & \multirow{2}{*}{Avg} & \multirow{2}{*}{std} & \multicolumn{3}{c}{Percentile} & \multirow{2}{*}{Total} \\ \cmidrule{4-6}
             & & & 25 & 50 & 75 & \\  \midrule
             c.p.s & 765.4 & 264.8 &  517 &  918 & 996 & 8,888,535 \\ 
             w.p.c & 62.8 & 148.0 & 15 & 26 & 58 & 558,327,230 \\ 
    \bottomrule
    \end{tabular}
    }
    \caption{Statistics of the \reddit{} dataset we analyse. c.p.s is content item (posts and comments) per subreddit, and w.p.c is word per content item.}
    \label{tab:data_statistics}
\end{table}

We download an image of \reddit{} posts and comments authored between January and August 2022\footnote{System limitations pose restrictions on the date range we can process -- an image of a single month is over 300GB.} using Pushshift's API.\footnote{\url{https://pushshift.io/}} We filter out posts and comments with fewer than ten words or with fewer than ten upvotes to reduce possible noise from low-quality text. We then merge the lists of posts and comments by subreddit, not further distinguishing between them for our analysis, and filter out subreddits that are tagged NSFW,\footnote{Not Safe For Work} have fewer than 5,000 subscribers, or fewer than 250 content examples. Finally, we down-sample large subreddits to 1,000 random samples due to computational constraints and remove non-English content. This process results in a dataset $\mathcal{D}$ of 11,616 unique subreddits. See \cref{tab:data_statistics} for dataset statistics.

\subsection{Value Extraction}
\label{sec:model}
For extracting Schwartz values and their polarity from text, we first extract the \textit{relevance} of a post or comment with each Schwartz value, then use a stance model to extract the \textit{polarity} of the sentence towards the relevant value. 
\paragraph{Value Relevance}
For extracting relevance, our approach is similar to ~\citet{van-der-meer-etal-2023-differences} in training a neural Schwartz Values relevance classifier. We use a DeBERTa model~\citep{he2021debertav3}, over RoBERTa, due its improved performance and speed. As a single post can potentially express several values simultaneously, the model is trained in a \textit{multi-label setting}, predicting a vector of 10 independent probabilities for each input.  We fine-tune the classifier on the concatenation of two supervised Schwartz values datasets, ValueNet \cite{qiu2022valuenet} and ValueArg \cite{kiesel-etal-2022-identifying}. Given a labelled triplet $(s, v, y)$, with $s$ being a string of text, $v$ the name of one of the ten Schwartz values (\Cref{tab:value_description}), and $y \in \{0, 1\}$, we construct the input $x = \specialtoken{[CLS]}v\specialtoken{[SEP]}s\specialtoken{[SEP]}$ and train the model to predict $p(y|x)$. Our classifier's performance on this dataset is similar to the figures reported by \citet{van-der-meer-etal-2023-differences}, namely a macro-averaged $F_1$ score of $0.76$ on the merged ValueArg and ValueNet datasets. Similar to \citet{van-der-meer-etal-2023-differences}, we collapse the two non-neutral labels (-1 and 1) in ValueNet into a single positive class.\footnote{We do this to follow the format of the ValueArg dataset, which only contains annotations for value relevance.} This model is trained, therefore, to predict the \textit{presence} of a value $v$ within a string $s$. 

\paragraph{Value Stance}

Prior work on value extraction neglects to model the polarity of content towards values, substantially limiting the insights one can draw. To detect the polarity of a comment or a post $s$ towards a value $v$, we train a separate stance model on ValueNet's non-neutral labels. Specifically, given a triplet $(s, v, y)$ where $v$ is a value expressed by $s$ and $y \in \{-1, 1\}$, we fit the stance model to predict $p_\text{stance}(y|x)$, where $x$ is constructed as described above. We refer readers to \cref{app:extractor_training_details} for additional training details about the classifiers. 

When extracting values of comments or posts from Reddit, we first use the relevance model to predict the probabilities for all Schwartz values for a given input instance, resulting in a vector $\schwartzvec{}_\text{rel} \in [0, 1]^{10}$. Specifically, given a string $s$, we construct $\schwartzvec{}_\text{rel} = [p(y^1|x^1), ..., p(y^{10}|x^{10})]$ by replacing $v^\evk$ in the construction of $x^\evk = \specialtoken{[CLS]}v^\evk \specialtoken{[SEP]}s\specialtoken{[SEP]}$ with each one of the ten possible Schwartz values. That is, each entry in the vector is the independent probability that $s$ expresses the value $v^\evk$, supporting a multi-label approach. Then, for each $k$ with $\schwartzvec{}_\text{rel}^\evk > 0.5$ (i.e., input text $s$ expresses the value $v^\evk$ with a probability greater than 0.5), we predict $p_\text{stance}(y^\evk|x^\evk)$, (i.e., the polarity of input text $s$ towards the value $v^\evk$). Thus, we construct a vector of probabilities $\schwartzvec{}_\text{stance}$ of dimensionality $10$, where each entry $\schwartzvec{}_\text{stance}^\evk$ is either in $[-1, 1]$ or Null (if $s$ does not express the value $v^\evk$ greater than 0.5). \Cref{tab:posts_examples} in \Cref{app:additional_material_examples} contains a sample of Reddit content and the associated Value Extractor model predictions.


%


\subsection{Evaluation of Value Extraction}
\label{sec:eval_value_model}

\begin{table}
    \centering
    \resizebox{0.5\textwidth}{!}{%
    \fontsize{10}{10}\selectfont
    \setlength{\tabcolsep}{4pt}
    \sisetup{table-format = 3.2, group-minimum-digits=3}
        \begin{tabular}{lrrrrrrrrrr}
        \toprule
          & AC &    BE &    CO &    HE &    PO &   SE &   SD &    ST &   TR &   UN \\
        \midrule
        Spearman $\rho$ & 0.73 & 0.66 & 0.66 & 0.6 & 0.56 & 0.8 & 0.46 & 0.5 & 0.6 & 0.73 \\
        NDCG@1 & 0.89 & 0.81 & 0.93 & 0.96 & 0.76 & 0.93 & 0.74 & 0.86 & 0.96 & 0.84 \\
        \bottomrule
    \end{tabular}
    }
\caption{Average Spearman's $\rho$ and NDCG@1 between the three annotators of the relevance model, per-value breakdown. AC=achievement, BE=benevolence, CO=conformity, HE=hedonism, PO=power, SE=security, SD=self-direction, ST=stimulation, TR=tradition, UN=universalism.
}
\label{tab:per_value_agreement_relevancy}
\end{table}

\begin{table}
    \centering
    \resizebox{0.5\textwidth}{!}{%
    \fontsize{10}{10}\selectfont
    \setlength{\tabcolsep}{4pt}
    \sisetup{table-format = 3.2, group-minimum-digits=3}
        \begin{tabular}{rrrrrrrrrr}
        \toprule
          AC &    BE &    CO &    HE &    PO &   SE &   SD &    ST &   TR &   UN \\
        \midrule
        0.51 & 0.57 & 0.61 & 0.77 & 0.30 & 0.45 & -0.27 & 0.77 & 0.67 & 0.26 \\
        \bottomrule
    \end{tabular}
    }
\caption{Cohen's Kappa between the two annotators of the stance model, per-value breakdown.
}

\label{tab:per_value_agreement_stance}
\end{table}

An important limitation of \citet{van-der-meer-etal-2023-differences} is the lack of direct evaluation of the value extraction model. The need for this becomes particularly pronounced considering the large domain shift of applying the model---which, similarly to us, was trained on debating data---to analyze content on platforms like \reddit{}. Therefore, we conduct a thorough evaluation to test the model's capabilities of extracting values from Reddit content. 

Ideally, we would want to assess the model's performance using a large, annotated dataset of randomly sampled Reddit posts. However, this is impractical because randomly selected posts are unlikely to contain any Schwartz values. Consequently, annotating these posts would yield a dataset with only a few positive examples. To address this challenge, we evaluate the Value Extraction models directly.

\paragraph{Relevance Model Evaluation}

We evaluate the relevance model by first using it to label 10,000 posts and 10,000 comments for predicting the presence of values. Thereafter, for each value $v$, we sample three posts or comments: one with high model confidence for the presence of the value (above 0.8), one with medium confidence (0.4-0.6), and one with low confidence (below 0.2). Three annotators then rank these comments/posts based on which are more related to value $v$, regardless of polarity. We repeat this process five times per value, totalling 50 rankings per annotator. See~\Cref{app:annotation_guidelines} for annotation guidelines and dataset samples. 

For annotator agreement, we calculated averaged Spearman's $\rho$, which looks at correlation amongst ranks. The agreement we found was 0.63, which is reasonable for such a subjective task.\footnote{\citet{kiesel-etal-2022-identifying} found the averaged agreement amongst annotators for value annotation to be $\alpha=0.49$.} Certain values (e.g., \schwartzvalue{security}, \schwartzvalue{universalism}) showed better agreement than others (e.g., \schwartzvalue{self-direction}). A full breakdown is available in \Cref{tab:per_value_agreement_relevancy}. For assessing model performance, we assigned a gold pseudo-ranking to each sample by averaging the annotators' rankings. We evaluate the relevance model's performance by comparing the model’s predicted ranking to this gold standard. The average Spearman's $\rho$ we obtain is $0.51$ (again, with values such as \schwartzvalue{security} outperforming other values such as \schwartzvalue{self-direction}). Looking at the top ranked content, the NDCG@1 we obtain is an impressive $0.87$, demonstrating that high certainty predictions made by our classifier are highly relevant to their corresponding value. With this, we can extract comments or posts relevant to certain values from the larger set.
See \Cref{app:additional_material_annotation} for examples of model misclassification and value breakdown.


\paragraph{Stance Model Evaluation}

We first sample 20 posts or comments per value $v$ where the relevance model predicted the presence of $v$ with high confidence. Two annotators then label the samples for stance, choosing between positive, negative, or neutral/unrelated (i.e., where no stance is clearly expressed or the relevance model misclassified the sample). The annotators achieved a Cohen's Kappa score of 0.47, representing moderate agreement, highlighting the challenging nature of the task (see \cref{tab:per_value_agreement_stance} for a per-value breakdown). However, post discussion on the disagreements, the annotators were able to converge on decisions for each sample, thus obtaining gold labels for the 200 samples. Finally, we apply the stance model to predict the stances of all positive or negative samples. The model predicts the stance towards values with an $F_1$ score of $0.72$. Given the subjective nature of the task, where one has to predict stance towards an abstract concept like human values and one where annotators often disagreed, we believe the model performs reasonably well. To highlight some of the trickier cases where the model is failing, \Cref{tab:stance_misclassification} in \Cref{app:additional_material_annotation} contains examples of model misclassification and breakdown into values. 


\begin{table}
    \centering
        \resizebox{0.5\textwidth}{!}{%
        \fontsize{10}{10}\selectfont
        \setlength{\tabcolsep}{4pt}
        \sisetup{table-format = 3.2, group-minimum-digits=3}
        \begin{tabular}{lR{1.5cm}R{1.5cm}R{1cm}R{1cm}}
        \toprule
        Value & Relevance mean & Relevance std & Stance mean & Stance std \\ 
        \midrule
        Tradition & 0.11 & 0.03 & -0.00 & 0.04 \\
        Benevolence & 0.30 & 0.12 & 0.01 & 0.29 \\
        Conformity & 0.23 & 0.06 & -0.17 & 0.32 \\
        Hedonism & 0.21 & 0.08 & 0.45 & 0.36 \\
        Power & 0.17 & 0.05 & 0.04 & 0.16 \\
        Achievement & 0.26 & 0.09 & 0.47 & 0.30 \\
        Self-direction & 0.19 & 0.09 & 0.25 & 0.31 \\
        Universalism & 0.43 & 0.10 & 0.39 & 0.27 \\
        Stimulation & 0.23 & 0.08 & 0.36 & 0.36 \\
        Security & 0.21 & 0.06 & 0.06 & 0.18 \\
    \bottomrule
    \end{tabular}
    }
\caption{Global statistics of the ten Schwartz values across the entire subreddit dataset.}
\label{tab:global_statistsics}
\end{table}

\subsection{Assigning Values to Subreddits}
\label{sec:assigning values}

We assign each subreddit $\mathsubreddit{} \in \mathcal{D}$ a single vector of Schwartz probabilities $\schwartzvec{}_\text{rel}(\mathsubreddit{})$ and a single vector of stances $\schwartzvec{}_\text{stance}(\mathsubreddit{})$. Given $\mathsubreddit{}$ with content ${(c_\evi)| \evi \in \mathsubreddit{}}$, where $c_\evi$ is either a post or a comment, we predict $\schwartzvec{}_\text{rel}(c_\evi)$ and $\schwartzvec{}_\text{stance}(c_\evi)$ from $c_\evi$ using the process above. 
Finally, we calculate $\schwartzvec{}_\text{rel}(\mathsubreddit{})$ by averaging over the predicted vectors $\schwartzvec{}_\text{rel}(c_\evi)$:
$$
    \schwartzvec{}_\text{rel}(\mathsubreddit{}) = \frac{1}{|\mathsubreddit{}|} \sum_{i \in \mathsubreddit{}} \schwartzvec{}_\text{rel}(c_\evi)
$$
We construct $\schwartzvec{}_\text{stance}(\mathsubreddit{})$ similarly, considering only non-Null entries. That is, each entry in $\schwartzvec{}_\text{stance}^\evk(\mathsubreddit{})$ is computed as
$$
    \schwartzvec{}_\text{stance}^\evk(\mathsubreddit{}) = 
    \frac{1}{|\mathsubreddit{}^\evk|} \sum_{i\in \mathsubreddit{}^\evk} \schwartzvec{}_\text{stance}^\evk(c_\evi)\text{,}
$$
where $\mathsubreddit{}^\evk = \{i \in \mathsubreddit{}| \schwartzvec{}_\text{stance}^\evk(c_\evi) \ne \text{Null}\}$.

\cref{tab:global_statistsics} details global statistics for the predicted relevance and stance for each Schwartz value across the entire subreddit dataset.

\section{Experiments}
\label{sec:experiments}
Our experiments serve two primary goals: first, to validate our approach by comparing the values extracted using our method to existing knowledge. Second, to demonstrate the extensibility of our method to new topics that have not been previously investigated in the social sciences. 
We conduct qualitative and quantitative evaluations of our approach in \Cref{sec:poc} and \Cref{sec:community_values}, to validate our method. Then, to uncover interesting phenomena and demonstrate the utility of our method, we investigate subreddits with differing opinions on controversial topics and compare our findings with existing studies.  Finally, in \Cref{sec:countries}, we correlate values extracted using our method to values gained through traditional approaches like surveys and questionnaires. 

\subsection{Qualitative Analysis}
\label{sec:poc}


    
 


    

We start by assessing how well our values extraction model performs on our \reddit{} dataset. 

\noindent \textbf{High relevance} Intuitively, certain Schwartz values are expected to be distinctly present in specific communities; e.g. \schwartzvalue{tradition} in religion-related communities. Therefore, we find subreddits with particularly strong signals of individual Schwartz values. For each value, we sort the subreddits' values probability vectors $\schwartzvec{}_{\text{rel}}(\mathsubreddit{})$ by their entry corresponding to the particular value. \cref{tab:strongest_signal_head} lists a sample of the top subreddits for each value; \cref{tab:strongest_signal} in \Cref{app:additional_material} lists the top 20 subreddits per value. Many of the subreddits collected for each value seem to be intuitively related to it (e.g., \subreddit{resumes} with \schwartzvalue{achievement}, \subreddit{conservation} with \schwartzvalue{Universalism}). The results are encouraging, demonstrating the effectiveness of our approach and its potential for conducting interesting analyses.

\noindent \textbf{Strong stances} Equivalently, we can also investigate which subreddits express the strongest \textit{stances} towards each value. For each value, we sort the subreddits' stance vectors $\schwartzvec{}_{\text{stance}}(\mathsubreddit{})$ by their entry corresponding to the particular value. \cref{tab:strongest_stances} in \Cref{app:additional_material} enumerates the 10 subreddits with the highest \textit{positive} and  \textit{negative} stance towards each value. Again, while some of the listed subreddits are intuitive (e.g., \subreddit{migraine} having a negative ``hedonism'' stance and \subreddit{raisingkids} having a strong positive stance towards ``achievement''), other subreddits are more surprising (\subreddit{TheHague} with a positive stance towards ``hedonism'').

\noindent \textbf{Value magnitude} We hypothesise that online communities differ not only in the set of values they express but also in the total \newterm{magnitude} of expressed values. 
Online communities pertaining to more objective topics (e.g., linear algebra) should express fewer Schwartz values than communities dedicated to subjective or controversial topics (e.g., politics).
To test this, for each subreddit $\mathsubreddit{} \in \mathcal{D}$ we calculate $\text{mag}(\mathsubreddit{}) = |\schwartzvec{}_{\text{rel}}(\mathsubreddit{})|_2$, the total magnitude of values expressed in the subreddit. The 20 subreddits with the highest and lowest value magnitudes are detailed in \cref{tab:top_magnitude} in \Cref{app:additional_material}. Subreddits with particularly high magnitudes are those communities that focus on debating, discussion, or emotional narratives. Examples of these include \subreddit{changemyview}, \subreddit{Adoption}, and \subreddit{PoliticalDiscussion}. Conversely, subreddits with notably low magnitudes are generally more objective and neutral in tone. These include \subreddit{crystalgrowing} and \subreddit{whatisthisfish}. Some subreddits that exhibit low-value magnitudes are dedicated to sharing photos taken by community members, e.g., \subreddit{astrophotography}. 

\subsection{Quantitative Analysis}
\label{sec:community_values}

We hypothesise that similar subreddit communities share similar values, and systematically investigate this using empirical evidence. We define three measures of similarity between subreddits: 

    \textbf{Value similarly}. We define $\similarityFunction{val}(\mathsubreddit{}_1, \mathsubreddit{}_2) = \text{cos}\left(\schwartzvec{}_\text{rel}(\mathsubreddit{}_1)||\schwartzvec{}_\text{stance}(\mathsubreddit{}_1), \schwartzvec{}_\text{rel}(\mathsubreddit{}_2)||\schwartzvec{}_\text{stance}(\mathsubreddit{}_2)\right)$, the cosine similarity between the concatenation of the relevance and stance vectors of the two subreddits.\footnote{We experimented with other formulations for calculating $\similarityFunction{val}(\mathsubreddit{}_1, \mathsubreddit{}_2)$, and arrived to similar results.} 
        
    \textbf{Semantic similarity}. We scrape the \newterm{public description} of each subreddit from its page (see \cref{fig:public_description} in \Cref{app:additional_material}), and embed these natural language descriptions using a sentence transformer\footnote{\url{https://www.sbert.net/}. We use the \texttt{all-mpnet-base-v2} pretrained model.} to construct a semantic embedding vector $\ve_\mathsubreddit{}$ for each subreddit $\mathsubreddit{}$\footnote{A limitation of this approach is that, occasionally, the public description is rather vague. However, a manual analysis we conducted reveals that this is rare.}. We now define $\similarityFunction{sem}(\mathsubreddit{}_1, \mathsubreddit{}_2) = \text{cos}(\ve_\mathsubreddit{}_1, \ve_\mathsubreddit{}_2)$. 
    
    \textbf{User similarity} For each pair of subreddits, we define their user similarity to be the \textit{overlap coefficient} between the users of the two subreddits. That is,
    $$
    \similarityFunction{usr}(\mathsubreddit{}_1, \mathsubreddit{}_2) = \frac{|{U}(\mathsubreddit{}_1) \cap {U}(\mathsubreddit{}_2)|}{\min(|{U}(\mathsubreddit{}_1)|, |{U}(\mathsubreddit{}_2)|)}
    $$
    Where $U(\mathsubreddit{})$ is the set of the subreddit's members.\footnote{As Reddit does not make this information publicly available, we estimate $U(\mathsubreddit{})$ by defining it as the set of users that had posted or commented in the subreddit within our dataset.}

To answer the question of if similar subreddits share similar values, we correlate $\similarityFunction{val}$ with $\similarityFunction{sem}$ and $\similarityFunction{usr}$ as follows. First, for each subreddit $\mathsubreddit{}$ we find 
$
\mostSimilarSubreddit{sem} = \text{argmax}_{\mathsubreddit{}'} \left[ \similarityFunction{sem}(\mathsubreddit{}, \mathsubreddit{}')\right]
$ 
and
$
\mostSimilarSubreddit{com} = \text{argmax}_{\mathsubreddit{}'} \left[ \similarityFunction{usr}(\mathsubreddit{}, \mathsubreddit{}')\right]
$.
If similar subreddits share a similar set of values, we should expect $\mathbb{E}_{\mathsubreddit{}} [\similarityFunction{val}(\mathsubreddit{}, \mostSimilarSubreddit{sem})]$ and $\mathbb{E}_{\mathsubreddit{}} [\similarityFunction{val}(\mathsubreddit{}, \mostSimilarSubreddit{com})]$ to be significantly larger than $\mathbb{E} \left[\similarityFunction{val}(\cdot,\cdot) \right]$. That is, the Schwartz values of two similar subreddits should be significantly closer to each other than the Schwartz values of two random subreddits. Our results, computed using empirical expectations over our dataset, confirm this hypothesis. Significant differences exist between the expected Schwartz values of similar and random subreddits. The expected values for both semantic and user similarity is $0.81$ versus $0.64$ for random subreddits. The z-test scores are $73.2$ and $74.4$, respectively, indicating a high level of statistical significance.


\begin{figure*}[ht]
  \centering
  \subfloat[Feminism]{\includegraphics[width=0.33\textwidth]{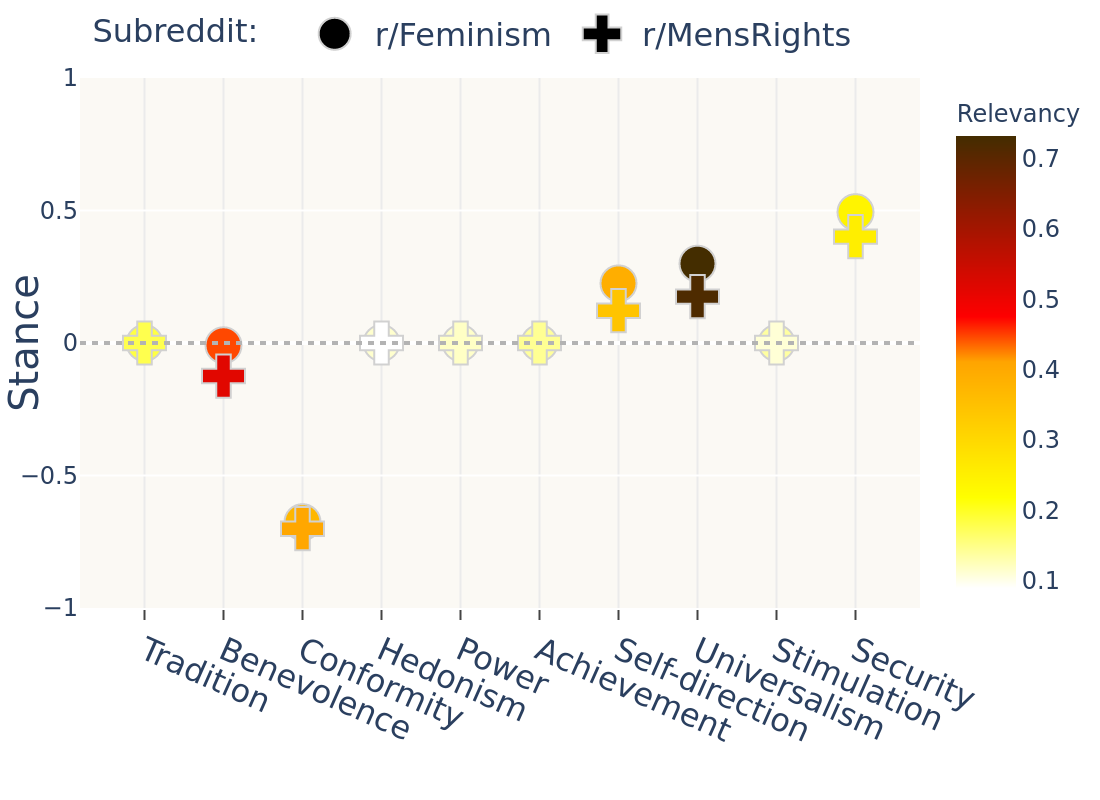} \label{fig:feminism}}
  \hspace{-2mm} 
  \subfloat[Atheism, Spirituality, and Religion]{\includegraphics[width=0.33\textwidth]{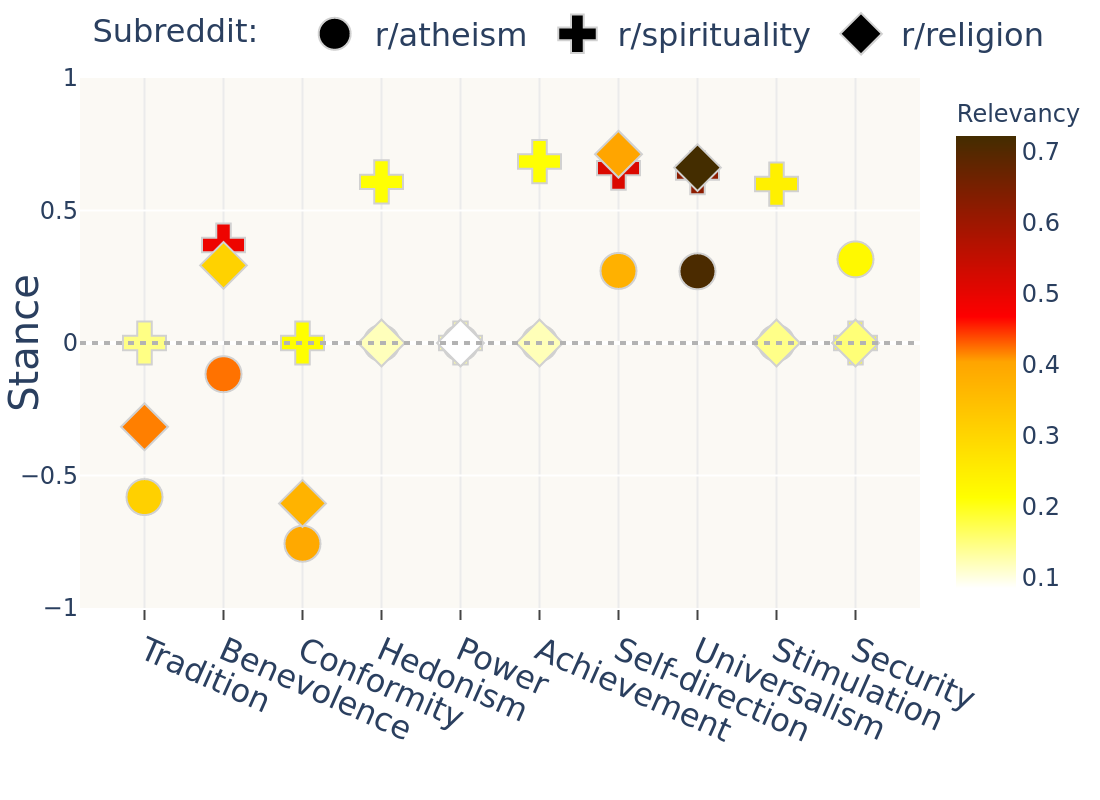} \label{fig:atheism}}
  \hspace{-2mm} 
  \subfloat[Veganism]{\includegraphics[width=0.33\textwidth]{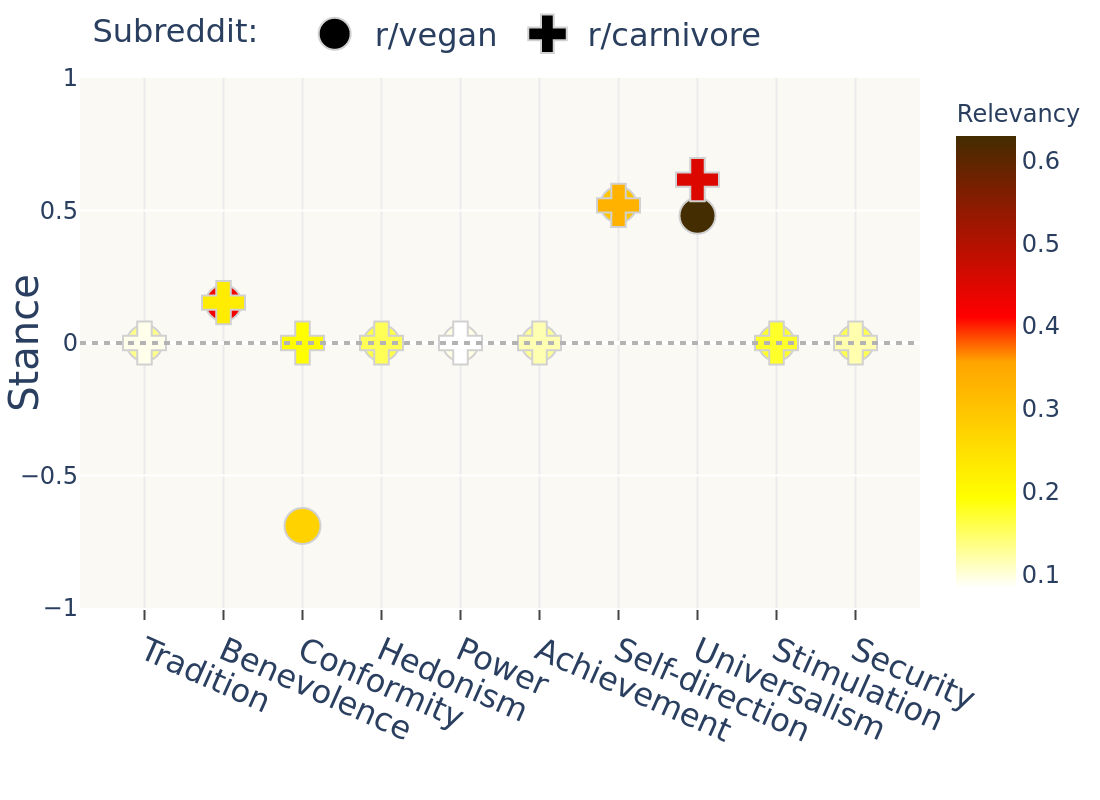} \label{fig:vegan}}\\
  \subfloat[Generations]{\includegraphics[width=0.33\textwidth]{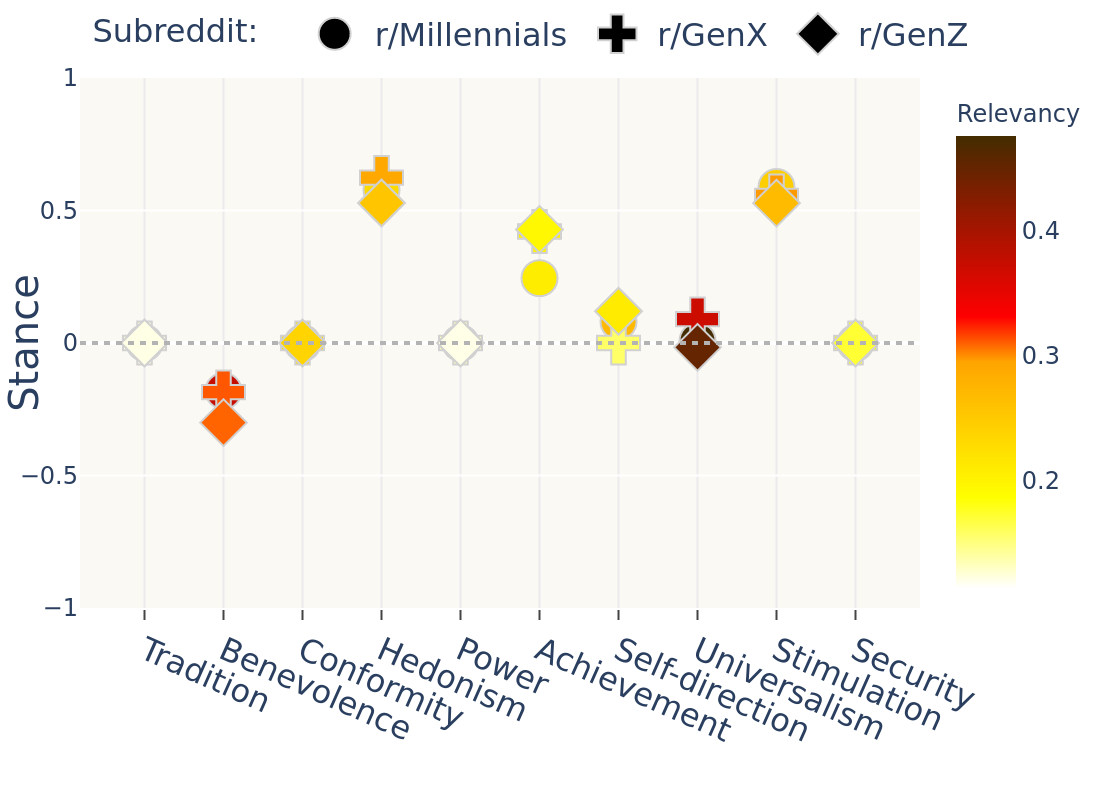} \label{fig:millennials}}
  \hspace{-2mm} 
  \subfloat[Communism vs Capitalism]{\includegraphics[width=0.33\textwidth]{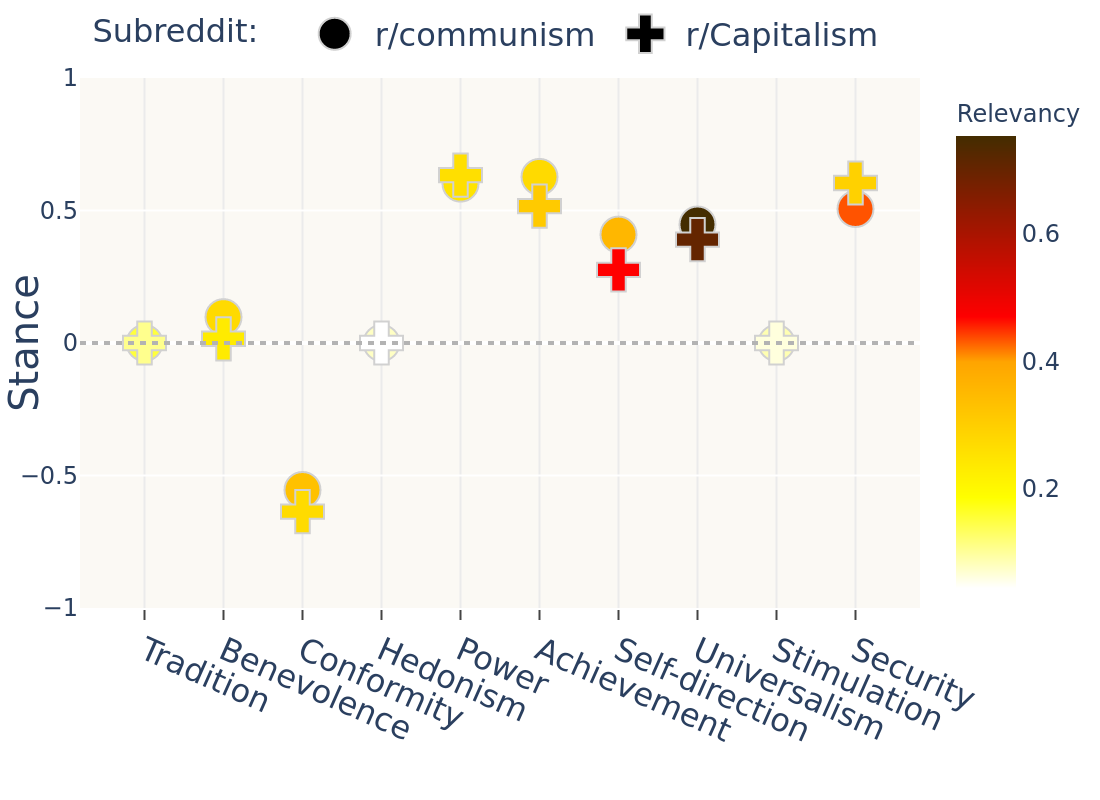} \label{fig:communism}}
  \hspace{-2mm} 
  \subfloat[Monarchism]{\includegraphics[width=0.33\textwidth]{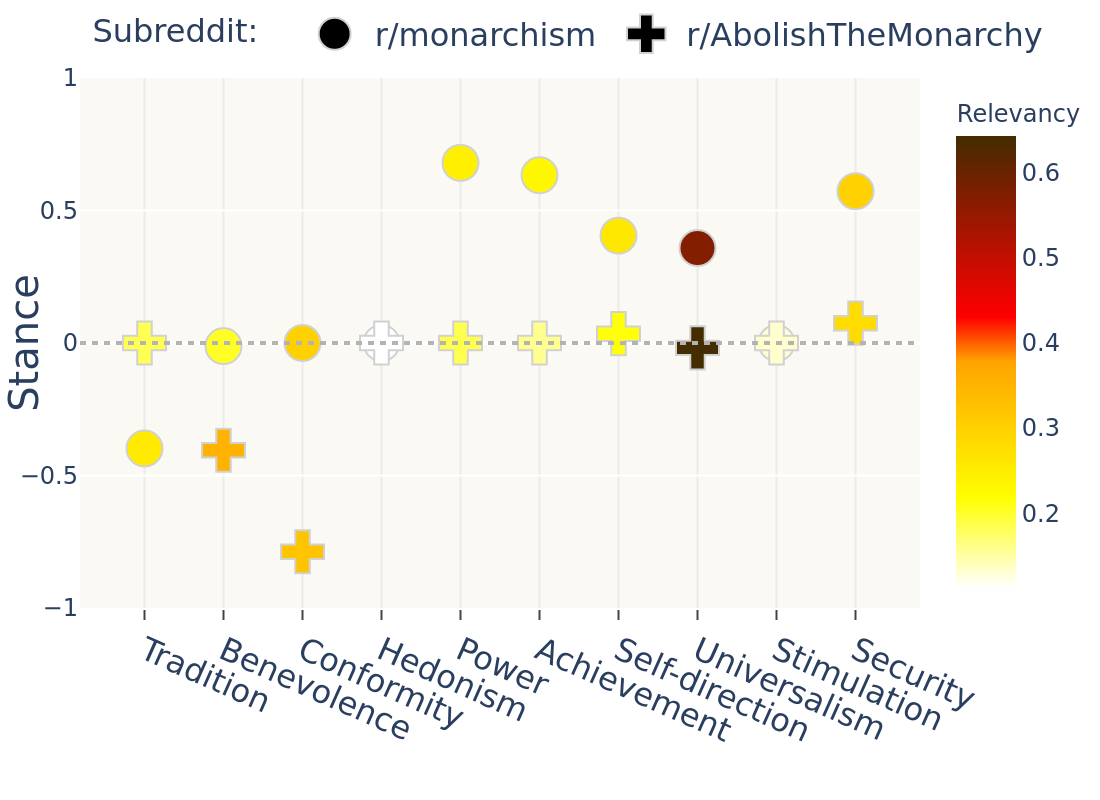} \label{fig:monarchism}}
  \caption{Radar plots displaying the ten Schwartz values of subreddits dedicated to controversial topics.}
  \label{fig:controversial_values}
\end{figure*}

\subsection{Controversial Topics}
\label{sec:controversial}
%
%
%
%
Different online communities have differing viewpoints on widely discussed issues based on their values.
To test whether we can also observe this in subreddits, we extend the analysis to \newterm{controversial topics}, based on Wikipedia's list of controversial topics.\footnote{\url{https://en.wikipedia.org/wiki/Wikipedia:List_of_controversial_issues}} 
For a selected topic, we identify subreddits with differing viewpoints on the topic, e.g., \subreddit{Communism} and \subreddit{Capitalism}. 
%
%
%

%
To establish the utility of our method and dataset, we analyse both topics studied by prior work in the social sciences, as well as previously unexplored topics that could be further studied. 
%
An overview of the results on all controversial topics can be found in Figure~\ref{fig:controversial_values}.
An overview of closely related subreddits (in terms of values) to each of the subreddits discussed here can be found in Appendix~\ref{app:controversial}.

\paragraph{Feminism}
%
%
In our investigation involving the Reddit communities of \subreddit{Feminism} and \subreddit{MensRights} (prior work \citep{Khan2020RedditMT, Witt_2020} characterized \subreddit{MensRights} as a non-feminist community based on the content shared by its members), we find both subreddits to be extremely anti-conformative (Figure~\ref{fig:feminism}). The values of \schwartzvalue{universalism}, \schwartzvalue{benevolence}, and \schwartzvalue{self-direction} were found to be highly relevant for both. \schwartzvalue{self-direction} as a value being more core to feminists, as compared to non-feminists has also been found in prior work exploring their values~\citep{zucker2010minding}.

\paragraph{Religion}
There are multiple large-scale studies aiming to establish the connection between religion and Schwartz values.
\citet{saroglou2004values} present a meta-analysis across 15 countries, finding that religious people prefer conservative values, e.g., \schwartzvalue{tradition}, and dislike values related to openness (\schwartzvalue{stimulation}, \schwartzvalue{self-direction}). 
We see similar trends in our comparison of \subreddit{atheism}, \subreddit{spirituality}, and \subreddit{religion} (Figure~\ref{fig:atheism}). 
%
%
While \subreddit{atheism} and \subreddit{religion} express \schwartzvalue{tradition}, interestingly, both subreddits have a negative stance towards it. Similarly, lower scores for \schwartzvalue{power}, \schwartzvalue{hedonism}, and \schwartzvalue{achievement} for those communities also align with previous findings. Interestingly, \subreddit{spirituality} has high relevance and a very positive stance towards some of those values, demonstrating the focus on the self.
%

%

\paragraph{Veganism}
Figure~\ref{fig:vegan} compares the values for \subreddit{vegan} and \subreddit{carnivore}. The largest difference can be observed in the value of \schwartzvalue{conformity}. \subreddit{Vegan} has a very negative stance towards it, representing them challenging the status quo.
\citet{holler2021differences} review studies related to the values of vegetarians/vegans and omnivores, and find that vegetarians were found to have a stronger relation to \schwartzvalue{universalism}. We find a similar pattern with \subreddit{vegan} having a slightly higher relevance but \subreddit{carnivore} having a more positive stance. They also found them to have a greater emphasis on \schwartzvalue{self-direction}, which aligns with our findings. 

\paragraph{Generations}
\citet{lyons2007empirical} study the differences in values across generations.
Similarly, we compare \subreddit{Millenials}, \subreddit{GenZ}, and \subreddit{GenX} in Figure~\ref{fig:millennials}. While the values of the different generations' subreddits are highly aligned, small differences can be observed, such as the more positive stance towards \schwartzvalue{achievement} in GenZ and GenX than Millenials. This contradicts \citet{lyons2007empirical}, which found Millenials to be more achievement focused than GenX.

\paragraph{Communism vs capitalism}
%
%
Here, we describe the difference in values for different economic ideologies, i.e., communism vs capitalism in Figure~\ref{fig:communism}. While \citet{schwartz1997influences} describe the effect of communism on Eastern Europe (e.g., high importance to conservatism and hierarchy values), they did not include the values held by the people supporting the ideology on a theoretical level. 
We find that contributors to \subreddit{communism} hold a more positive stance towards \schwartzvalue{achievement}, whereas contributors to \subreddit{Capitalism} hold high relevance values for \schwartzvalue{self-direction} and \schwartzvalue{security}. Both subreddits have high relevance with \schwartzvalue{universalism}. 
\paragraph{Monarchism}
We further present values on monarchism in Figure~\ref{fig:monarchism}, as another example of a controversial topic yet to be studied in terms of values. 
We find contributors to \subreddit{AbolishTheMonarchy} to be against conformity, with a negative stance towards \schwartzvalue{benevolence}.
The contributors to \subreddit{monarchism} have a high relevance to \schwartzvalue{tradition}, but a negative stance towards it. They also converse a lot more positively about \schwartzvalue{power}, \schwartzvalue{achievement}, and \schwartzvalue{security}.


\subsection{Correlation with Surveys}
\label{sec:countries}

\begin{figure*}[ht]
    \centering
    \includegraphics[width=0.999\linewidth]{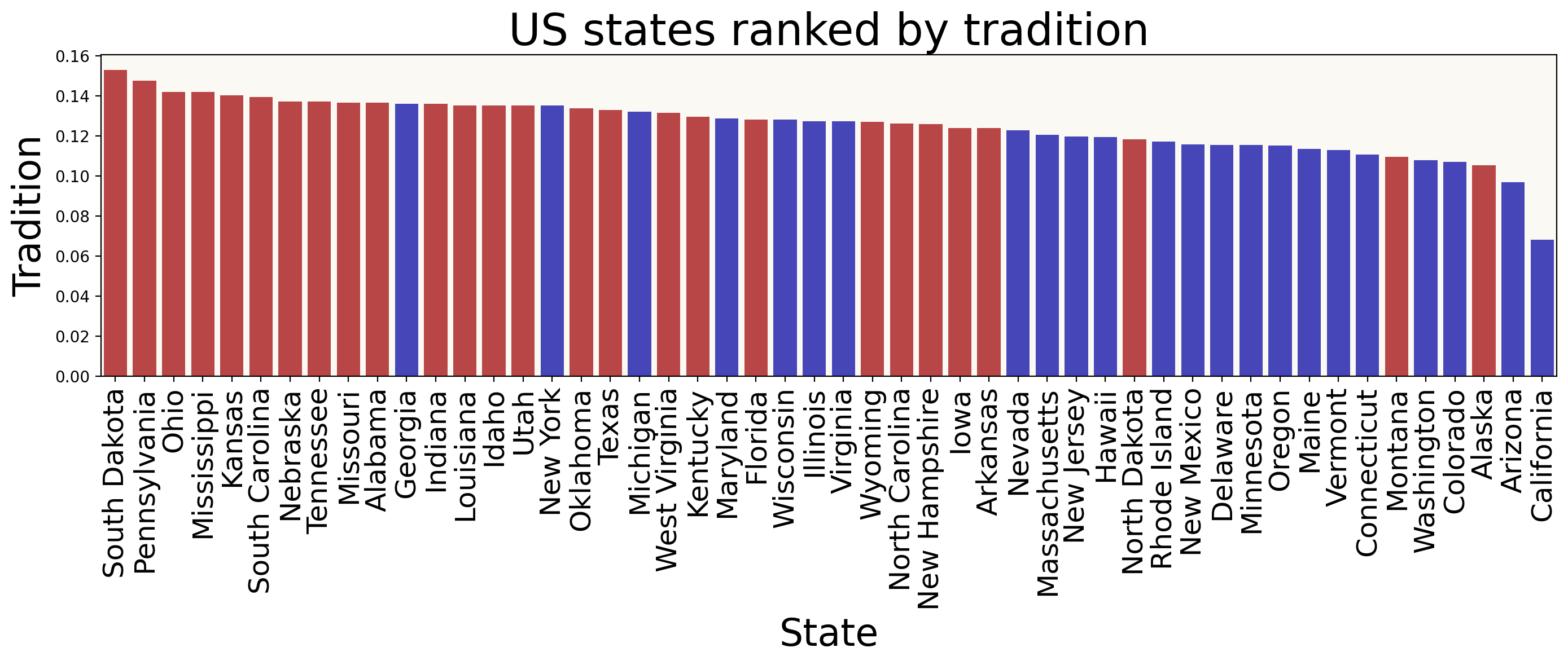}
    \caption{US states sorted by their \schwartzvalue{tradition} values extracted from Reddit, colour coded based on the 2020 US election results (Blue -- democratic majority. Red -- republican majority).}
    \label{fig:state}
\end{figure*}

Finally, to understand how well aligned values expressed in online communities are with values of real-world communities, we compare the Schwartz values extracted from \reddit{} with those obtained from traditional questionnaire methods. 

\noindent \textbf{US States} First, we use our method to investigate the premise that conservative US states are more traditional than liberal states. We correlate the \schwartzvalue{tradition} relevance value extracted using our method from the states' subreddits with responses to a survey on conservative ideologies across U.S. states \cite{PewProject_2015}, finding a Spearman's $R$ of $0.55$ (p-value  $ < 0.0001$). Moreover, when correlating these values with a survey on state religiosity levels \cite{PewProject_2014}, we find a Spearman's $R$ of $0.63$ (p-value $< 0.0001$).~\cref{fig:state} displays the \schwartzvalue{tradition} value of the 50 states subreddits, colour-coded by the 2020 US election results -- ``red'' states clearly tend to have higher \schwartzvalue{tradition} values than ``blue'' states.

\noindent \textbf{Country-level values} Next, we extend this correlation experiment to additional Schwartz values and countries other than the USA. \citet{schwartz2022measuring} used questionnaires to analyse the Schwartz values associated with 49 countries,\footnote{While the survey is conducted with various populations and regions, here we use the term ``countries'' for simplicity.} which we correlate with values extracted from subreddits related to these countries using our classifier. We start by manually identifying a set of subreddits relevant to each of the 49 countries.\footnote{We searched for country-related subreddits online and filtered based on their public description and front-page posts, disregarding ones that tourists or non-residents majorly use.} We successfully found at least one subreddit for 41 countries. We also determined which language is the most likely to be used online by citizens of the country (and not, for example, by tourists).\footnote{We acknowledge that this is a simplification, as in many countries, more than one language is commonly spoken.} See \cref{tab:countries_data} in \Cref{app:additional_material} for a comprehensive list of countries and their associated languages and subreddits. Next, we collected posts written in the country's language\footnote{We used \texttt{lingua} (\url{https://github.com/pemistahl/lingua-py}) to detect the language of each post.} from the country's subreddits. We then randomly sampled 2,000 posts and 2,000 comments per country, excluding countries with fewer than 250 total samples. \Cref{listing:countries} lists the 32 countries that passed this filter. Finally, we used Google Translate API to translate the posts into English\footnote{We manually scanned a sample of the translations and deemed them as having satisfactory quality.} and applied our approach to extract Schwartz values for each country.


The above-mentioned process generates a country-value matrix of dimensions $32 \times 10$ of Schwartz values obtained from \reddit{}. We compare this matrix to the $49 \times 10$ the country-value described in \citet{schwartz2022measuring} by removing rows corresponding to countries that do not exist in \Cref{listing:countries}. We then correlate the columns of the two now-identical matrices using Spearman's $\rho$. In line with previous studies \citep{nasif1991methodological, joseph-etal-2021-mis}, we find no correlation between the values extracted from SM and values extracted from questionnaires---an average Spearman's $\rho$ of $-0.03$. \cref{tab:values_correlation} in \Cref{app:additional_material_experimetns} reports the results in full. This lack of correlation may arise from the distinct demographics of Reddit users compared to questionnaire respondents or the unique dynamics of online interaction, which tends to be more reactionary compared to survey responses~\citep{joseph-etal-2021-mis}. Our results confirm findings from previous studies that dynamics of online behaviour differs from offline, and should not directly be used as a proxy without further qualitative investigation.

\section{Conclusion}
\label{sec:conclusion}
In this paper, we apply Schwartz's Theory of Human Values to Reddit at scale, offering scholars a complementary tool for studying online communities. We train 
 and thoroughly evaluate supervised value extraction models to detect the presence and polarity towards human values expressed in language used on social media. Using these, we conduct extensive analysis of nine million posts across 12k popular subreddits. Through our analysis, we both confirm previous findings from the social sciences, as well as uncover novel insights into the values expressed within various online communities, shedding light on existing patterns and outliers that warrant further investigation.



\section*{Limitations}
\label{sec:limitations}
The main limitations of our work stem from the nature of the task itself. Given the inherent subjectivity and complexity of assigning human values to texts, a certain level of noise---aleatoric uncertainty---is unavoidable. This unpredictability can be amplified by epistemic uncertainty, which arises from the limitations of the models we trained. Although we have thoroughly validated the labels generated by the model (\Cref{sec:eval_value_model}), the predictions can sometimes be noisy or inconsistent.

Another limitation of our approach is the consolidation of the values of posts in a subreddit into a single vector. While this is necessary for understanding the values of communities, it discounts the values the individual posts of users that constitute these communities. This also opens up much room for future research, particularly in studying the internal dynamics of online communities and understanding the role that values play at the individual post level. 

Lastly, while our approach can be leveraged to identify phenomena worthy of further investigation, it lacks the means to explain these observations fully. Values are an abstract concept that is challenging to quantify and analyse, with the values frameworks themselves drawing criticism~\citep{jackson2020legacy}. We argue that interdisciplinary collaboration with experts in psychology and sociology is essential to understand these phenomena and their implications properly. Such collaborations will not only enrich our understanding of online communities but also contribute to developing more robust and nuanced machine learning models in the future, given the inclusion of such text in the training data.

\section*{Broader Impact and Ethical Considerations}
\label{sec:ethics}
We believe our work goes beyond the methodological toolkit of social scientists. The accessibility, large scale, and relatively high quality of Reddit data have positioned it as a valuable resource for training data for diverse NLP tasks \citep{overbay-etal-2023-mredditsum,blombach-etal-2020-corpus,huryn-etal-2022-automatic}. It opens the door to training models that are 
better attuned to the specific needs of various communities, potentially protected ones. Recognizing the values of online communities, often utilized as training data for LM development, is crucial for making informed decisions about incorporating the data and contributing to the creation of models that genuinely reflect the diverse perspectives within these communities~\citep{arora-etal-2023-probing}. These also percolate into downstream applications of LMs~\citep{jackesch-2023-cowriting} and have a broader impact on their use.

As for ethical considerations, we made specific efforts to ensure that our work does not impair the privacy and anonymity of Reddit users \cite{sugiura2017ethical}. We refrain from attributing values to individual users and instead study communities as a whole by aggregating and condensing individual data points into a single vector. Nevertheless, many online communities were created to serve as a safe space for vulnerable individuals, where they share highly sensitive and private information.  Therefore, research on Reddit and other online communities should make utmost efforts to respect these spaces and handle their data with care.




\section*{Acknowledgements}
This research was co-funded by a DFF Sapere Aude research leader grant under grant agreement No 0171-00034B and a DFF Research Project 1 under grant agreement No 9130-00092B, and supported by the Pioneer Centre for AI, DNRF grant number P1.

\bibliography{anthology,custom}

\clearpage

\appendix
\section{Annotation Guidelines}
\label{app:annotation_guidelines}

\begin{lstlisting}[label=listing:guidelines, caption=Guidelines for the task of annotating the dataset used to evaluate the relevancy model., numbers=none]

    For each three consecutive rows in the table (colour-coded), start by reading the three sentences in column B. Next, review the Schwartz value associated with the three sentences (listed in column C) and ensure you understand its meaning, using the figure at the top of the table as a reference. Then, order the sentences according to the extent to which they express the value, ignoring the stance of the sentences towards the value. In other words, our focus is solely on whether the value is expressed, not on how it is expressed. Even if a sentence violates the value, it still expresses it. For instance, the sentence ``I never went to church'' expresses the value of ``Tradition,'' despite contradicting it.


    This task is subjective, and in many cases, there is no single correct answer. If you are uncertain about a particular instance, respond to the best of your ability.
\end{lstlisting}

\begin{lstlisting}[label=listing:guidelines_stance, caption=Guidelines for the task of annotating the dataset used to evaluate the stance model., numbers=none]

    For each row in the table, start by reading the sentence in column A. Next, review the Schwartz value associated with the sentence in column B and ensure you understand its meaning, using the figure at the top of the table as a reference. Then, determine the stance of the sentence towards the value. If the sentence supports the value (i.e., the value is expressed in a positive way, or the sentence describes a situation where the value is maintained), select ``positive''. If the sentence violates the value (i.e., the value is expressed negatively, or the sentence describes a situation where the value is disregarded or broken), select ``negative''. If the sentence simply does not express the value, select ``N\A''. For instance, the sentence ``I never went to church'' has a negative stance towards the value of ``Tradition''. Conversly, the sentence ``family comes first for her'' has a positive stance towards the value.


    This task is subjective, and in many cases, there is no single correct answer. If you are uncertain about a particular instance, respond to the best of your ability.
\end{lstlisting}

\begin{figure*}[t!]
    \centering
    \includegraphics[width=\textwidth]{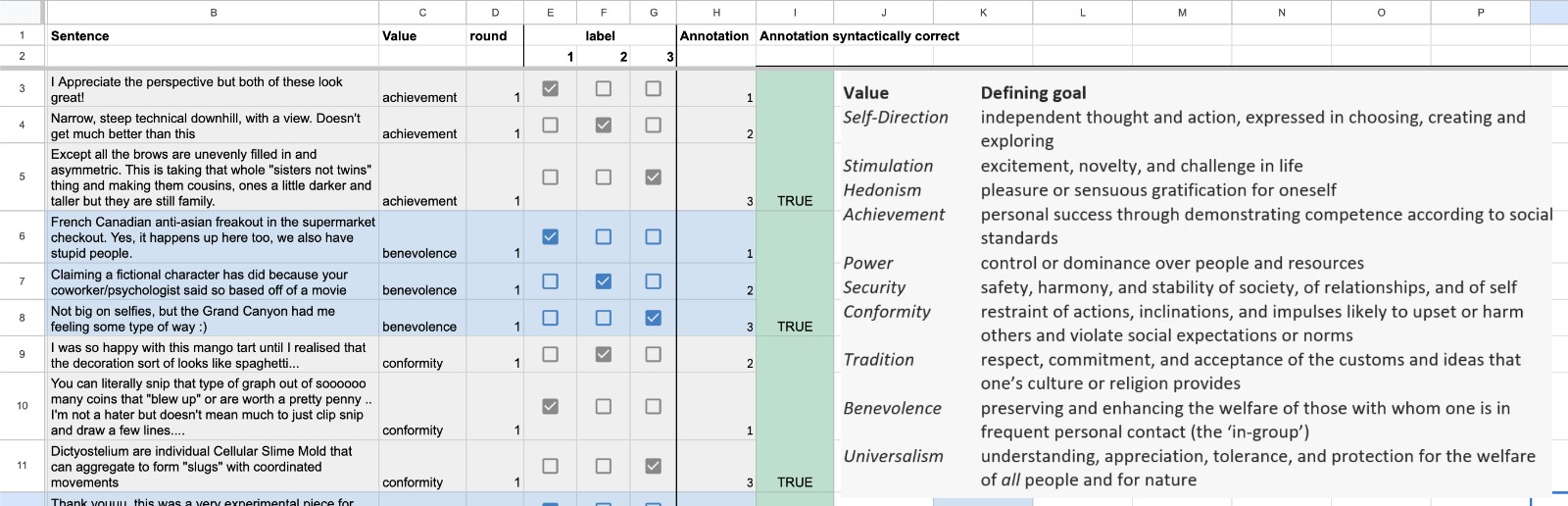}
    \caption{Print screen of the annotation task of the relevance model. The three annotators were tasked with ranking triplets of sentences according to the extent to which they express the value of column C.}
    \label{fig:annotation_job}
\end{figure*}

\begin{figure*}[t!]
    \centering
    \includegraphics[width=\textwidth]{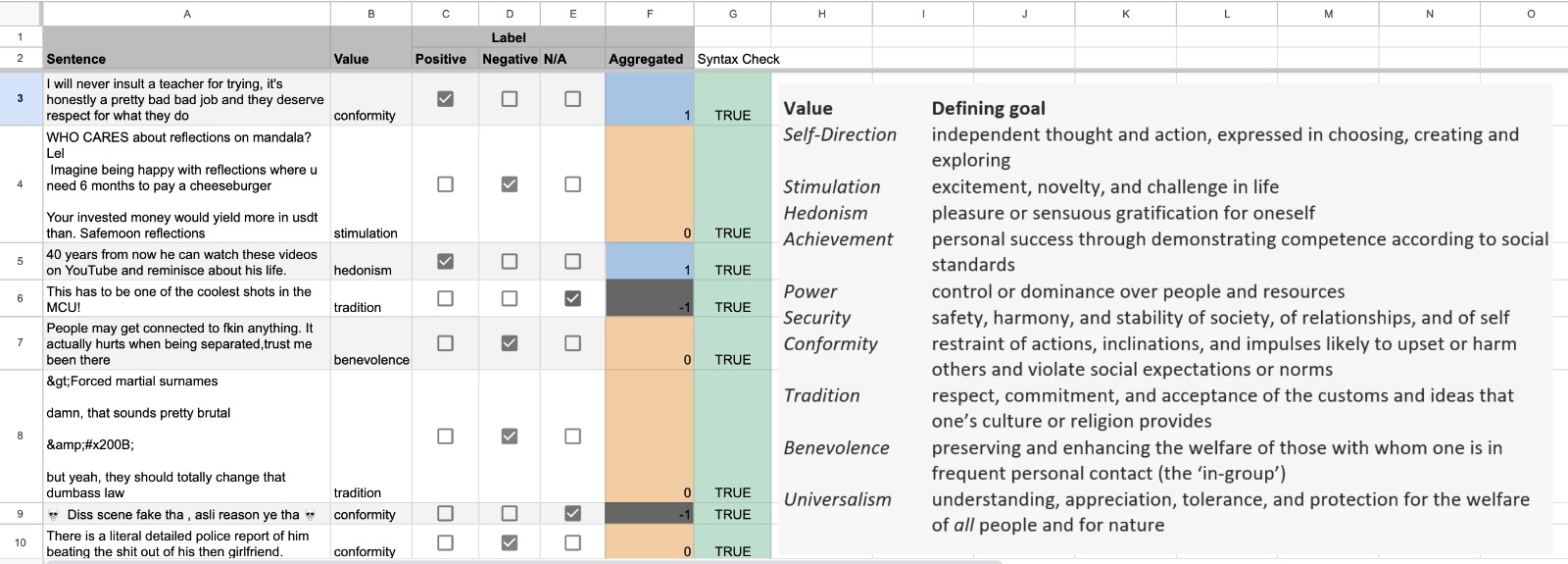}
    \caption{Print screen of the annotation task of the stance model. Two annotators were tasked to determine the stance of the sentence in column A towards the value in column B.}
    \label{fig:annotation_job_stance}
\end{figure*}

As described in \Cref{sec:eval_value_model}, three authors of this paper annotated Reddit content to evaluate the Value Extraction Models. To evaluate the relevance model, each annotator ranked triplets $(s_1, s_2, s_3)$ according to the extent each sentence $s_i$ expresses a given value $v$. In total, each annotator ranked 50 triplets, or 5 triplets per value. \Cref{fig:annotation_job} contains a print screen of the annotation task, whereas \Cref{listing:guidelines} specifies the annotation guidelines. 

To evaluate the stance model, each annotator was given a sentence $s$ and a value $v$, and was tasked to determine the stance of $s$ towards $v$ (one of ``positive'', ``negative'', or ``irrelevant/neutral'', indicating that sentence $s$ does not express the value $v$). Each annotator annotated a total of 200 sentences, or 20 per value. \Cref{fig:annotation_job_stance} contains a print screen of the annotation task, whereas \Cref{listing:guidelines_stance} specifies the annotation guidelines.

\section{Additional Material}
\label{app:additional_material}

\subsection{Dataset Statistics}
\label{app:dataset_statistics}

\label{sec:data}


\subsection{Annotation and Evaluation}
\label{app:additional_material_annotation}

\begin{table}
    \centering
    \resizebox{0.5\textwidth}{!}{%
    \fontsize{10}{10}\selectfont
    \setlength{\tabcolsep}{4pt}
    \sisetup{table-format = 3.2, group-minimum-digits=3}
        \begin{tabular}{rrrrrrrrrr}
        \toprule
        AC &    BE &    CO &    HE &    PO &   SE &   SD &    ST &   TR &   UN \\
        \midrule
         -0.16 & -0.31 & -0.04 & 0.04 & -0.21 & 0.10 & -0.12 & 0.16 & 0.22 & 0.01 \\
        \bottomrule
    \end{tabular}
    }
\caption{Spearman correlation between the Schwartz values obtained from a questionnaire and the values extracted from \reddit{}.
}
\label{tab:values_correlation}
\end{table}




\begin{table}

 \centering
 \resizebox{1.0\columnwidth}{!}
 {%
 \fontsize{8}{8}\selectfont
 \sisetup{table-format = 3.2, group-minimum-digits=3}
 
 \begin{tabular}{lp{4cm}p{1.5cm}}
 \toprule
 \textbf{Val} & \textbf{Sentence}  & \textbf{Annotated Average Rank} \\ 
 \midrule
SD & idk it brings some discussions to the community, can`t be bad & 2.33 \\
PO & You really gave it all you had very impressive content  & 2.66 \\
CO & Turkish guy debates Marxism and Freud with a street upper & 2.33 \\ 
BE & Claiming a fictional character has did because your coworker/psychologist said so based off of a movie & 2.33 \\
PO & What is each of your favorite song from Mercurial World to perform live? & 2.66 \\
 \bottomrule
 \end{tabular}
}

\caption{Instances where the relevance model predicted that the sentence expresses a value with high confidence (above 0.8), whereas the annotators concluded that the sentence does not express the value.}
\label{tab:rel_misclassification}
\end{table}

\begin{table}

 \centering
 \resizebox{1.0\columnwidth}{!}
 {%
 \fontsize{8}{8}\selectfont
 \sisetup{table-format = 3.2, group-minimum-digits=3}
 
 \begin{tabular}{lp{4cm}p{1.5cm}}
 \toprule
 \textbf{Val} & \textbf{Sentence}  & \textbf{Predicted (wrong) Stance} \\ 
 \midrule
PO & Crypto traders using the war to make money. This is a new low & Pos \\
HE & God forbid you pay someone for their content that you enjoy.  & Neg \\
CO &SLPT: drive at speed in reverse gear through speed cameras to get fines repaid by the police & Pos \\ 
SD & If the island keeps drifting, the UK would be part of the US again? & Neg \\
UN &The double standards of western media towards third world countries & Pos \\
SD & The coffee is probably making it worse. People forget it’s a drug with consequences. & Neg \\
 \bottomrule
 \end{tabular}
}

\caption{Instances where the stance model misclassified a sentence.}
\label{tab:stance_misclassification}
\end{table}


\Cref{tab:rel_misclassification} and \Cref{tab:stance_misclassification} present cases where the relevance model and stance models, respectively, misclassified instances.

\subsection{Prediction Examples}
\label{app:additional_material_examples}
\begin{table*}

 \centering
 \resizebox{1.0\textwidth}{!}
 {%
 \fontsize{12}{12}\selectfont
 \sisetup{table-format = 3.2, group-minimum-digits=3}
 
 \begin{tabular}{p{12cm}cc}
 \toprule
 \textbf{Sentence}  & \textbf{Predicted Value} & \textbf{Predicted Stance} \\ 
 \midrule
The green party was literally removed from the ballot in key battleground states in the 2020 election. & Power & Negative \\
When you’re trained by a leader you become a leader. & Power & Positive \\
I've been learning jazz for about 8 months and still can't write a single good peace, is that normal?Just wondering if Im not deaf because every time I'm trying to write something I end up raging. In general I make music for 4-5 years if that matters. & Achievement & Negative \\
My first ever attempt at finger crochet (or any crochet for that matter)! I’m really proud of myself and just wanted to share somewhere :) & Achievement & Positive \\
I have never had a birthday celebrationSo my parents are Jehovah's witnesses I'm not but I have never received a birthday present and this year I'm really depressed I want to celebrate but I'm broke & Hedonism & Negative \\
To love more is Beautiful.I wish you Wonderful evening and happy weekend & Hedonism & Positive \\
This video gave me anxiety. It’s so inefficient with its movements that it made me squirm. & Stimulation & Negative \\
This awe inspiring film is the culmination of 3 months working underwater, diving, exploring and filming the most remote corners of the Great Barrier Reef. It showcases some of the most stunning scenes, schooling sharks and most vibrant coral reef the ocean has to offer. Hope you enjoy! & Stimulation & Positive \\
Having to lie so that you don't have to work Saturdays is funny and wholesome smh & Self-direction & Negative \\
If you want to be better at squatting, this is fine.  If you are going for a specific aesthetic, evaluate/critique your physique for any lagging parts areas, then choose the exercises that help bring them up. & Self-direction & Positive \\
"The only moral abortion is MY abortion" or "Grifters gonna Grift" & Universalism & Negative \\
I REALLY THINK ALL OF US HUMANS NEED TO BE REPRESENTED PROPERLY, REGARDLESS OF WHAT LANGUAGE PACKS WE HAVE INSTALLED. & Universalism & Positive \\
And then blaming the teacher when half the class fails. & Benevolence & Negative \\
please ban mentions of hasan off of LSF. his mental health is being affected by the harassment from this sub.	& Benevolence & Positive \\
Respect my religion while I shove it down your throat.... & Tradition & Negative \\
Korobeiniki: known in the West as "Tetris music"; in Russian, it\'s a folk song about courtship. & Tradition & Positive \\
There is a literal detailed police report of him beating the shit out of his then girlfriend.	& Conformity & Negative \\
Indian minister drinks dirty water from 'holy' river polluted with sewage to show locals it's safe... ends up in hospital days later & Conformity & Positive \\
This is in Mexico. The guy can literally pay the cops to look the other way, and he obviously has the money to do so & Security & Negative \\
India also built the largest border walls with pakistan, that too in the Himalayan foothills and thar desert. Really helped in curbing cross border terror though. & Security & Positive \\

 \bottomrule
 \end{tabular}
}
\caption{Examples of posts and comments from Reddit and the values and stances that the relevance and stance models predicted for each instance.}
\label{tab:posts_examples}
\end{table*}

\Cref{tab:posts_examples} contains examples of posts and comments from Reddit and the values and stances that the relevance and stance models predicted for each instance.

\subsection{Experiments}
\label{app:additional_material_experimetns}

\begin{table*}[ht]
    \centering
    \resizebox{0.98\textwidth}{!}{%
    \fontsize{8}{8}\selectfont
    \sisetup{table-format = 3.2, group-minimum-digits=3}
\begin{tabular}{lp{14cm}}
\toprule
Value & Subreddits \\
\midrule
Tradition      &                                                                                                  \subredditnegative{0.42igion (0.42)}, \subreddit{Anglicanism (0.42)}, \subreddit{TraditionalCatholics (0.37)}, \subreddit{AskAChristian (0.37)}, \subredditnegative{AcademicBiblical (0.36)}, \subreddit{Catholic (0.36)}, \subreddit{AskBibleScholars (0.36)}, \subreddit{Episcopalian (0.36)}, \subredditpositive{AskAPriest (0.35)}, \subredditnegative{Antitheism (0.34)}, \subreddit{OrthodoxChristianity (0.34)}, \subreddit{Catholicism (0.33)}, \subredditnegative{antitheistcheesecake (0.33)}, \subredditnegative{agnostic (0.32)}, \subredditnegative{atheistmemes (0.32)}, \subreddit{CatholicMemes (0.32)}, \subredditpositive{Bible (0.32)}, \subredditnegative{RadicalChristianity (0.31)}, \subredditnegative{atheism (0.31)}, \subredditnegative{0.31igiousfruitcake (0.31)} \\ \midrule
Benevolence    &                                                                                                                                            \subredditpositive{coparenting (0.87)}, \subreddit{Stepmom (0.85)}, \subreddit{BipolarSOs (0.84)}, \subreddit{theotherwoman (0.84)}, \subreddit{stepparents (0.84)}, \subreddit{AdultChildren (0.83)}, \subreddit{ADHD\_partners (0.83)}, \subredditpositive{CaregiverSupport (0.83)}, \subreddit{SingleParents (0.83)}, \subredditpositive{Petloss (0.83)}, \subredditpositive{Adoption (0.83)}, \subreddit{emotionalabuse (0.83)}, \subreddit{EstrangedAdultChild (0.82)}, \subredditpositive{AgingParents (0.82)}, \subreddit{FriendshipAdvice (0.82)}, \subreddit{AnxiousAttachment (0.82)}, \subreddit{Adopted (0.82)}, \subreddit{attachment\_theory (0.82)}, \subreddit{AsOneAfterInfidelity (0.82)}, \subreddit{NarcAbuseAndDivorce (0.82)} \\ \midrule
Conformity     &                                       \subredditnegative{AgainstHateSubreddits (0.54)}, \subredditnegative{reclassified (0.52)}, \subredditnegative{TheseFuckingAccounts (0.50)}, \subredditnegative{AmIFreeToGo (0.50)}, \subredditnegative{ModSupport (0.50)}, \subredditnegative{modsbeingdicks (0.49)}, \subredditnegative{ModsAreKillingReddit (0.49)}, \subredditnegative{policebrutality (0.49)}, \subredditnegative{modhelp (0.48)}, \subredditnegative{WatchRedditDie (0.47)}, \subredditnegative{Bad\_Cop\_No\_Donut (0.47)}, \subredditnegative{Chiraqhits (0.47)}, \subredditnegative{JustUnsubbed (0.46)}, \subredditnegative{legaladviceofftopic (0.46)}, \subredditnegative{circlebroke2 (0.46)}, \subreddit{HOA (0.46)}, \subredditnegative{BadNeighbors (0.46)}, \subredditnegative{redditrequest (0.46)}, \subredditnegative{law (0.46)}, \subredditnegative{fuckHOA (0.46)} \\ \midrule
Hedonism       &                                                          \subredditpositive{crossdressing (0.56)}, \subredditpositive{transadorable (0.56)}, \subredditpositive{Autumn (0.55)}, \subredditpositive{TheMidnight (0.53)}, \subredditpositive{dykesgonemild (0.53)}, \subredditpositive{cakedecorating (0.53)}, \subredditpositive{FreeCompliments (0.53)}, \subredditpositive{GothStyle (0.52)}, \subredditpositive{NailArt (0.52)}, \subredditpositive{cozy (0.51)}, \subredditpositive{PlusSizeFashion (0.51)}, \subredditpositive{MTFSelfieTrain (0.50)}, \subredditpositive{oldhagfashion (0.50)}, \subredditpositive{femboy (0.50)}, \subredditpositive{transpositive (0.50)}, \subredditpositive{christmas (0.49)}, \subredditpositive{malepolish (0.49)}, \subredditpositive{RainbowEverything (0.49)}, \subredditpositive{gaybrosgonemild (0.49)}, \subredditpositive{HappyTrees (0.49)} \\ \midrule
Power          &                                                                           \subredditpositive{FundRise (0.41)}, \subredditpositive{AskEconomics (0.38)}, \subredditpositive{finance (0.37)}, \subredditpositive{acorns (0.37)}, \subredditpositive{HistoricalWhatIf (0.37)}, \subredditpositive{geopolitics (0.37)}, \subredditpositive{strongblock (0.36)}, \subredditpositive{UWMCShareholders (0.36)}, \subredditpositive{debtfree (0.36)}, \subredditpositive{qyldgang (0.36)}, \subredditpositive{rocketpool (0.36)}, \subredditpositive{Anchor (0.36)}, \subredditpositive{IndianStreetBets (0.36)}, \subredditpositive{Economics (0.35)}, \subredditpositive{dividends (0.35)}, \subredditpositive{VoteBlue (0.35)}, \subredditpositive{EuropeanFederalists (0.35)}, \subredditpositive{defi (0.35)}, \subredditpositive{ValueInvesting (0.35)}, \subredditpositive{AlibabaStock (0.35)} \\ \midrule
Achievement    &                                                \subredditpositive{passive\_income (0.60)}, \subredditpositive{FundRise (0.60)}, \subredditpositive{ValueInvesting (0.59)}, \subredditpositive{OptionsMillionaire (0.59)}, \subredditpositive{EngineeringResumes (0.58)}, \subredditpositive{xboxachievements (0.57)}, \subredditpositive{algotrading (0.57)}, \subredditpositive{FuturesTrading (0.57)}, \subredditpositive{resumes (0.55)}, \subredditpositive{abmlstock (0.55)}, \subredditpositive{AllCryptoBets (0.55)}, \subredditpositive{cryptostreetbets (0.55)}, \subredditpositive{dividends (0.55)}, \subredditpositive{SatoshiBets (0.54)}, \subredditpositive{Forex (0.54)}, \subredditpositive{EANHLfranchise (0.54)}, \subredditpositive{IndianStockMarket (0.54)}, \subredditpositive{quant (0.54)}, \subredditpositive{Blogging (0.54)}, \subredditpositive{aabbstock (0.54)} \\ \midrule
Self-direction &                   \subredditpositive{CapitalismVSocialism (0.67)}, \subredditpositive{ScientificNutrition (0.67)}, \subredditpositive{Abortiondebate (0.66)}, \subredditpositive{DebateAnarchism (0.65)}, \subredditpositive{BasicIncome (0.63)}, \subredditpositive{DebateAVegan (0.62)}, \subredditpositive{Neuropsychology (0.62)}, \subredditpositive{changemyview (0.61)}, \subredditpositive{AskEconomics (0.61)}, \subredditpositive{AskPsychiatry (0.59)}, \subredditpositive{AskDID (0.59)}, \subredditpositive{nutrition (0.58)}, \subredditpositive{AskSocialScience (0.58)}, \subredditpositive{healthcare (0.58)}, \subredditpositive{askpsychology (0.58)}, \subredditpositive{radicalmentalhealth (0.58)}, \subredditpositive{ketoscience (0.57)}, \subredditpositive{DebateReligion (0.57)}, \subredditpositive{Marxism (0.57)}, \subredditpositive{PsychedelicTherapy (0.57)} \\ \midrule
Universalism   &  \subredditpositive{DebateAVegan (0.88)}, \subredditpositive{IsraelPalestine (0.87)}, \subredditpositive{DebateEvolution (0.85)}, \subredditpositive{DebateAnarchism (0.85)}, \subredditpositive{DebateReligion (0.84)}, \subredditpositive{changemyview (0.84)}, \subredditpositive{Abortiondebate (0.84)}, \subredditpositive{CapitalismVSocialism (0.84)}, \subredditpositive{AskSocialScience (0.84)}, \subreddit{EndlessWar (0.83)}, \subredditpositive{LeftWingMaleAdvocates (0.83)}, \subredditpositive{DebateAnAtheist (0.83)}, \subredditpositive{TrueUnpopularOpinion (0.82)}, \subredditpositive{PoliticalDiscussion (0.82)}, \subredditpositive{Ask\_Politics (0.82)}, \subredditpositive{IntellectualDarkWeb (0.82)}, \subredditpositive{chomsky (0.82)}, \subredditpositive{AskFeminists (0.81)}, \subredditpositive{Marxism (0.81)}, \subredditpositive{DebateCommunism (0.81)} \\ \midrule
Stimulation    &                                                                              \subredditpositive{Autumn (0.51)}, \subredditpositive{crossdressing (0.51)}, \subredditpositive{TheMidnight (0.51)}, \subredditpositive{transadorable (0.51)}, \subredditpositive{FreeCompliments (0.49)}, \subredditpositive{dykesgonemild (0.48)}, \subredditpositive{cakedecorating (0.48)}, \subredditpositive{xboxachievements (0.47)}, \subredditpositive{PlusSizeFashion (0.47)}, \subredditpositive{GothStyle (0.47)}, \subredditpositive{cozy (0.46)}, \subredditpositive{MTFSelfieTrain (0.46)}, \subredditpositive{christmas (0.46)}, \subredditpositive{AltJ (0.46)}, \subredditpositive{Madonna (0.46)}, \subredditpositive{weddingdress (0.46)}, \subredditpositive{NailArt (0.46)}, \subredditpositive{transpositive (0.46)}, \subredditpositive{happy (0.46)}, \subredditpositive{Hobbies (0.46)} \\ \midrule
Security       &                       \subredditpositive{CredibleDefense (0.66)}, \subredditpositive{ww3 (0.65)}, \subredditpositive{syriancivilwar (0.62)}, \subredditpositive{geopolitics (0.62)}, \subredditpositive{EndlessWar (0.62)}, \subredditpositive{warinukraine (0.59)}, \subredditpositive{AfghanConflict (0.58)}, \subredditpositive{UkrainianConflict (0.58)}, \subredditpositive{UkraineConflict (0.56)}, \subredditpositive{CombatFootage (0.55)}, \subredditpositive{LessCredibleDefence (0.55)}, \subreddit{GunsAreCool (0.55)}, \subredditpositive{war (0.55)}, \subredditpositive{FutureWhatIf (0.55)}, \subredditpositive{nuclearweapons (0.54)}, \subredditpositive{WarCollege (0.53)}, \subredditpositive{UkraineInvasionVideos (0.53)}, \subredditpositive{UkraineRussiaReport (0.53)}, \subredditpositive{RussiaUkraineWar2022 (0.52)}, \subredditpositive{UkraineWarReports (0.52)} \\ \bottomrule

\end{tabular}
}
    \caption{Subreddits with the highest signal for each one of the ten Schwartz values (the number in parenthesis indicates the magnitude of the signal). The stance of \textcolor{OliveGreen}{Green} subreddits towards the value is positive (above $0.2$), whereas \textcolor{BrickRed}{Red} indicates negative stance (below $-0.2)$.}
    \label{tab:strongest_signal}
\end{table*}



\begin{table*}
    \centering
    \resizebox{0.98\textwidth}{!}{%
    \fontsize{8}{8}\selectfont
    \sisetup{table-format = 3.2, group-minimum-digits=3}

\begin{tabular}{lp{6cm}|p{6cm}}
\toprule
  Value & Positive Stance & Negative Stance \\ \midrule
Tradition & \subreddit{Ankrofficial}, \subreddit{lds}, \subreddit{CharliDamelioMommy}, \subreddit{Christian}, \subreddit{AskAPriest}, \subreddit{Bible}, \subreddit{bahai}, \subreddit{Quakers}, \subreddit{PrismaticLightChurch}, \subreddit{OrthodoxChristianity} & \subreddit{SuperModelIndia}, \subreddit{Jewdank}, \subreddit{EXHINDU}, \subreddit{DesiMeta}, \subreddit{linguisticshumor}, \subreddit{exmuslim}, \subreddit{AsABlackMan}, \subreddit{Satan}, \subreddit{IndoEuropean}, \subreddit{AfterTheEndFanFork} \\ \midrule
Benevolence & \subreddit{freebsd}, \subreddit{RandomKindness}, \subreddit{Terraform}, \subreddit{Petloss}, \subreddit{nextjs}, \subreddit{Wetshaving}, \subreddit{AllCryptoBets}, \subreddit{NixOS}, \subreddit{vancouverhiking}, \subreddit{ansible} & \subreddit{FromDuvalToDade}, \subreddit{CrimeInTheD}, \subreddit{NBAYoungboy}, \subreddit{40kOrkScience}, \subreddit{LILUZIVERTLEAKS}, \subreddit{DuvalCounty}, \subreddit{Phillyscoreboard}, \subreddit{Chiraqhits}, \subreddit{SummrsXo}, \subreddit{CARTILEAKS} \\ \midrule
Conformity & \subreddit{Ankrofficial}, \subreddit{nanotrade}, \subreddit{NervosNetwork}, \subreddit{Vechain}, \subreddit{steroids}, \subreddit{USCIS}, \subreddit{treelaw}, \subreddit{Stellar}, \subreddit{cancun}, \subreddit{JapanFinance} & \subreddit{Animewallpaper}, \subreddit{kencarson}, \subreddit{FromDuvalToDade}, \subreddit{LilDurk}, \subreddit{Cookierun}, \subreddit{freddiegibbs}, \subreddit{SummrsXo}, \subreddit{DestroyLonely}, \subreddit{Gunna}, \subreddit{okbuddydaylight} \\ \midrule
Hedonism & \subreddit{eastside}, \subreddit{RedditPHCyclingClub}, \subreddit{OaklandFood}, \subreddit{carcamping}, \subreddit{CryptoMars}, \subreddit{VeganBaking}, \subreddit{TheHague}, \subreddit{ZZZ\_Official}, \subreddit{pottedcats}, \subreddit{ambientmusic} & \subreddit{depression}, \subreddit{TIHI}, \subreddit{Shark\_Park}, \subreddit{willowbramley}, \subreddit{Phillyscoreboard}, \subreddit{migraine}, \subreddit{BodyDysmorphia}, \subreddit{SuicideWatch}, \subreddit{2meirl4meirl}, \subreddit{anhedonia} \\ \midrule
Power & \subreddit{Yotsubros}, \subreddit{Fitness}, \subreddit{CryptoMoonShots}, \subreddit{infertility}, \subreddit{AllCryptoBets}, \subreddit{ketoscience}, \subreddit{steroids}, \subreddit{ProgrammingLanguages}, \subreddit{cryptostreetbets}, \subreddit{cooperatives} & \subreddit{masterhacker}, \subreddit{Stake}, \subreddit{uknews}, \subreddit{RustConsole}, \subreddit{uspolitics}, \subreddit{OPBR}, \subreddit{BidenIsNotMyPresident}, \subreddit{capitalism\_in\_decay}, \subreddit{Patriot911}, \subreddit{occupywallstreet} \\ \midrule
Achievement & \subreddit{infertility}, \subreddit{theravada}, \subreddit{edrums}, \subreddit{raisingkids}, \subreddit{CryptoMars}, \subreddit{Yotsubros}, \subreddit{cozy}, \subreddit{gaidhlig}, \subreddit{PrismaticLightChurch}, \subreddit{learnrust} & \subreddit{DreamStanCringe}, \subreddit{FGOmemes}, \subreddit{AntiTrumpAlliance}, \subreddit{BidenWatch}, \subreddit{misanthropy}, \subreddit{Patriot911}, \subreddit{TRUTHsocialWatch}, \subreddit{Instagram}, \subreddit{TwitterCringe}, \subreddit{Negareddit} \\ \midrule
Self-direction & \subreddit{Mosses}, \subreddit{jungle}, \subreddit{LandscapingTips}, \subreddit{icecreamery}, \subreddit{esp32}, \subreddit{SatoshiBets}, \subreddit{rust\_gamedev}, \subreddit{openstreetmap}, \subreddit{QuantumComputing}, \subreddit{cryptostreetbets} & \subreddit{Negareddit}, \subreddit{misanthropy}, \subreddit{PeopleFuckingDying}, \subreddit{BoomersBeingFools}, \subreddit{AmericanFascism2020}, \subreddit{RepublicanValues}, \subreddit{FoxFiction}, \subreddit{ParlerWatch}, \subreddit{libsofreddit}, \subreddit{FragileWhiteRedditor} \\ \midrule
Universalism & \subreddit{SatoshiBets}, \subreddit{CryptoMars}, \subreddit{AllCryptoBets}, \subreddit{nextjs}, \subreddit{AskAstrophotography}, \subreddit{GardenWild}, \subreddit{vancouverhiking}, \subreddit{dungeondraft}, \subreddit{EatCheapAndVegan}, \subreddit{GraphicsProgramming} & \subreddit{CrimeInTheD}, \subreddit{FromDuvalToDade}, \subreddit{Phillyscoreboard}, \subreddit{DaDumbWay}, \subreddit{DuvalCounty}, \subreddit{Chiraqhits}, \subreddit{BruceDropEmOff}, \subreddit{punchableface}, \subreddit{ConservativeRap}, \subreddit{NYStateOfMind} \\ \midrule
Stimulation & \subreddit{CryptoMars}, \subreddit{cryptostreetbets}, \subreddit{JoshuaTree}, \subreddit{wonderdraft}, \subreddit{reenactors}, \subreddit{AllCryptoBets}, \subreddit{estoration}, \subreddit{yerbamate}, \subreddit{GiftIdeas}, \subreddit{OaklandFood} & \subreddit{depression}, \subreddit{Shark\_Park}, \subreddit{heck}, \subreddit{TwitterCringe}, \subreddit{depressed}, \subreddit{CommercialsIHate}, \subreddit{Demps}, \subreddit{D\_Demps}, \subreddit{christenwhitmansnark}, \subreddit{Sleepparalysis} \\ \midrule
Security & \subreddit{BoringCompany}, \subreddit{haskell}, \subreddit{ProgrammingLanguages}, \subreddit{rust}, \subreddit{psychoanalysis}, \subreddit{crypto}, \subreddit{steroids}, \subreddit{AskComputerScience}, \subreddit{infertility}, \subreddit{DebateEvolution} & \subreddit{holdmycosmo}, \subreddit{davidbowiecirclejerk}, \subreddit{RoastMyCat}, \subreddit{Chiraqhits}, \subreddit{DaDumbWay}, \subreddit{okbuddydaylight}, \subreddit{bottomgear}, \subreddit{SuicideWatch}, \subreddit{CrimeInTheD}, \subreddit{Bombing} \\ 
\bottomrule
\end{tabular}
}
\caption{Subreddits expressing each value's strongest positive and negative stances.}
\label{tab:strongest_stances}
\end{table*}

\begin{table}[ht]
    \centering
    
    \fontsize{10}{10}\selectfont
\begin{tabular}{lp{5cm}}
\toprule
 Magnitude & Subreddits \\ \midrule
 Maximal & \subreddit{DebateAnarchism}, \subreddit{Abortiondebate}, \subreddit{therapyabuse}, \subreddit{CapitalismVSocialism}, \subreddit{changemyview}, \subreddit{AvoidantAttachment}, \subreddit{LeftWingMaleAdvocates}, \subreddit{coparenting}, \subreddit{ADHD\_partners}, \subreddit{DebateAVegan}, \subreddit{Ask\_Politics}, \subreddit{IsraelPalestine}, \subreddit{PoliticalDiscussion}, \subreddit{AskSocialScience}, \subreddit{NarcAbuseAndDivorce}, \subreddit{AskDID}, \subreddit{attachment\_theory}, \subreddit{Adoption}, \subreddit{kpoprants}, \subreddit{TrueUnpopularOpinion}

\\ \midrule
 Minimal & \subreddit{vegan1200isplenty}, \subreddit{caloriecount}, \subreddit{Watchexchange}, \subreddit{Brogress}, \subreddit{crystalgrowing}, \subreddit{sneakermarket}, \subreddit{gundeals}, \subreddit{buildapcsales}, \subreddit{whatisit}, \subreddit{NMSCoordinateExchange}, \subreddit{BulkOrCut}, \subreddit{astrophotography}, \subreddit{legodeal}, \subreddit{whatisthisthing}, \subreddit{whatsthisfish}, \subreddit{filmfashion}, \subreddit{TipOfMyFork}, \subreddit{1500isplenty}, \subreddit{safe\_food}, \subreddit{Repbudgetfashion}
\\
\bottomrule

\end{tabular}
    \caption{Subreddits with the highest and lowest total magnitude of Schwartz values, calculated according to $\text{mag}(\mathsubreddit{}) = |\schwartzvec{}_{\text{rel}}(\mathsubreddit{})|_2$.}
    \label{tab:top_magnitude}
\end{table}

\begin{figure}[ht]
    \centering
    \includegraphics[width=0.95\linewidth]{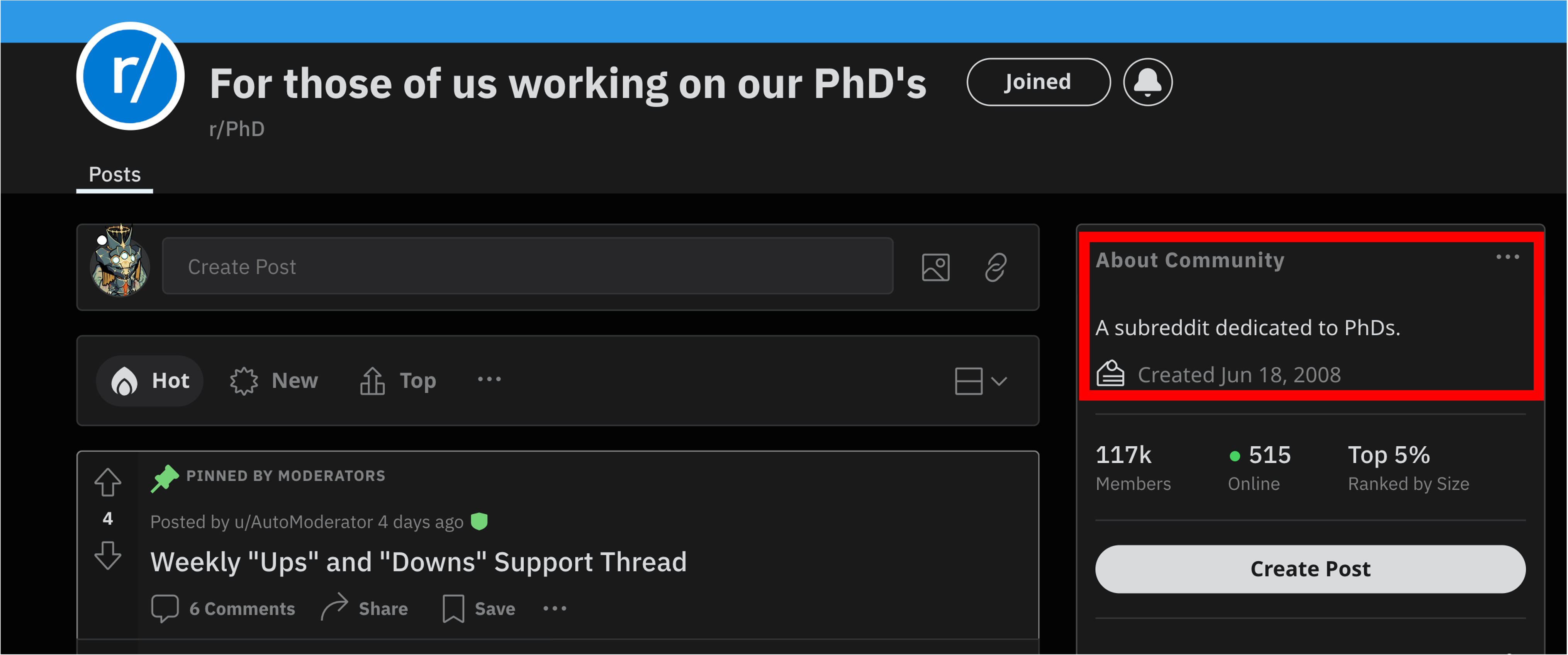}
    \caption{Location of the ``Public description'' attribute on a subreddit page.}
    \label{fig:public_description}
\end{figure}


\begin{table*}[ht]
    \centering
    \resizebox{0.98\textwidth}{!}{%
    \fontsize{10}{10}\selectfont
    \sisetup{table-format = 3.2, group-minimum-digits=3}

\begin{tabular}{lll}
\toprule
Country &        Language &  Subreddits \\ 
\midrule
Australia      &     English &                           \subreddit{australia}, \subreddit{Australia\_}, \subreddit{australian}, \subreddit{sydney}, \subreddit{melbourne} \\
Brazil         &  Portuguese &                            \subreddit{brasil}, \subreddit{brasilivre}, \subreddit{GTAorBrazil}, \subreddit{brasilia}, \subreddit{saopaulo} \\
China          &     Chinese &                                                                       \subreddit{China}, \subreddit{China\_irl}, \subreddit{real\_China\_irl} \\
Colombia       &     Spanish &                                                                       \subreddit{Colombia}, \subreddit{ColombiaReddit}, \subreddit{Bogota} \\
Costa Rica     &     Spanish &                                                                                                   \subreddit{costarica}, \subreddit{Ticos} \\
Croatia        &    Croatian &                                                                                                    \subreddit{croatia}, \subreddit{zagreb} \\
Czech Republic &       Czech &                                                                                                                          \subreddit{czech} \\
Ecuador        &     Spanish &                                                                                                                        \subreddit{ecuador} \\
Estonia        &    Estonian &                                                                                                                          \subreddit{Eesti} \\
Faroe Islands  &         All &                                                                                                                   \subreddit{Faroeislands} \\
Finland        &     Finnish &                                                                                                                          \subreddit{Suomi} \\
France         &      French &                                                                                                      \subreddit{france}, \subreddit{paris} \\
Georgia        &    Georgian &                                                                                                \subreddit{Sakartvelo}, \subreddit{tbilisi} \\
Germany        &      German &                                                              \subreddit{de}, \subreddit{deuchland}, \subreddit{Munich}, \subreddit{berlin} \\
Ghana          &     English &                                                                                                                          \subreddit{ghana} \\
Greece         &       Greek &                                                                                                                         \subreddit{greece} \\
Hong Kong      &     Chinese &                                                                                                                       \subreddit{HongKong} \\
Iceland        &   Icelandic &                                                                                                                        \subreddit{Iceland} \\
India          &         All &                                          \subreddit{india}, \subreddit{IndiaSpeaks}, \subreddit{unitedstatesofindia}, \subreddit{askindia} \\
Indonesia      &  Indonesian &                                                                                                                      \subreddit{indonesia} \\
Israel         &      Hebrew &                                                                             \subreddit{Israel}, \subreddit{telaviv}, \subreddit{jerusalem} \\
Italy          &     Italian &                                                                                                      \subreddit{italy}, \subreddit{Italia} \\
Japan          &    Japanese &                                                                                 \subreddit{japan}, \subreddit{tokyo}, \subreddit{newsokur} \\
New Zealand    &     English &                                                                                               \subreddit{newzealand}, \subreddit{auckland} \\
Oman           &      Arabic &                                                                                                                           \subreddit{Oman} \\
Peru           &     Spanish &                                                                                                                           \subreddit{PERU} \\
Philippines    &     English &                                                                                                                    \subreddit{Philippines} \\
Poland         &      Polish &                                                                                                                         \subreddit{Polska} \\
Portugal       &  Portuguese &                                                                                \subreddit{portugal}, \subreddit{lisboa}, \subreddit{porto} \\
Romania        &    Romanian &                                      \subreddit{Romania}, \subreddit{bucuresti}, \subreddit{iasi}, \subreddit{cluj}, \subreddit{timisoara} \\
Serbia         &     Serbian &                                                                                                                         \subreddit{serbia} \\
Slovakia       &      Slovak &                                                                                                                       \subreddit{Slovakia} \\
South Africa   &         All &                                                                    \subreddit{southafrica}, \subreddit{capetown}, \subreddit{johannesburg} \\
South Korea    &      Korean &                                                                                                                         \subreddit{hanguk} \\
Spain          &     Spanish &                                                           \subreddit{Espana}, \subreddit{spain}, \subreddit{Madrid}, \subreddit{Barcelona} \\
Sweden         &     Swedish &                                                  \subreddit{Sverige}, \subreddit{sweden}, \subreddit{stockholm}, \subreddit{svenskpolitik} \\
Turkey         &     Turkish &                                                            \subreddit{Turkey}, \subreddit{istanbul}, \subreddit{ankara}, \subreddit{Ismir} \\
Ukraine        &   Ukrainian &                                                                                                \subreddit{ukraine}, \subreddit{Ukraine\_UA} \\
United Kingdom &         All &  \subreddit{unitedkingdom}, \subreddit{london}, \subreddit{manchester}, \subreddit{CasualUK}, \subreddit{unitedkingdom}, \subreddit{askuk} \\
USA            &         All &                                                                                                                      \texttt{All states subreddits} \\
Vietnam        &  Vietnamese &                                                                                                                        \subreddit{VietNam} \\
\bottomrule
\end{tabular}
}
    \caption{Countries, languages and subreddits used in the correlation study.}
    \label{tab:countries_data}
\end{table*}

\begin{lstlisting}[label=listing:countries, caption=List of countries used in the correlation experiment in \Cref{sec:countries}., numbers=none]
    Turkey, South Africa, Estonia, Romania, USA, Costa Rica, Poland, New Zealand, Brazil, Greece, Finland, Ukraine, Croatia, Colombia, Slovakia, United Kingdom, Iceland, Czech Republic, Italy, Sweden, China, Ecuador, Germany, Peru, Indonesia, Serbia, Spain, India, Australia, Philippines, Portugal, France, 
\end{lstlisting}


\noindent \textbf{\Cref{tab:strongest_signal}} Subreddits with the highest signal for each one of the ten Schwartz values. The stance of \textcolor{OliveGreen}{Green} subreddits towards the value is positive (above $0.2$), whereas \textcolor{BrickRed}{Red} indicates negative stance (below $-0.2)$.

\noindent \textbf{\Cref{tab:strongest_stances}} Subreddits expressing the strongest positive and negative stances for each value.

\noindent \textbf{\Cref{tab:top_magnitude}} Subreddits with the highest and lowest total magnitude of Schwartz values, calculated according to $\text{mag}(\mathsubreddit{}) = |\schwartzvec{}_{\text{rel}}(\mathsubreddit{})|_2$.

\noindent \textbf{\Cref{fig:public_description}} Location of the ``Public description'' attribute on a subreddit page, used to calculate $\similarityFunction{sem}(\mathsubreddit{}_1, \mathsubreddit{}_2)$ in \Cref{sec:community_values}.



\noindent \textbf{\Cref{listing:countries}} List of countries used in the correlation experiment in \Cref{sec:countries}.

\subsection{Controversial Topics -- Extended}\label{app:controversial}
Additionally to the analysis presented in Section~\ref{sec:controversial}, Table~\ref{tab:app:controversial} displays the ten most similar subreddits in terms of values. We use value similarity as described in Section~\ref{sec:community_values} to determine the closest subreddits to each of the controversial topic subreddits.
\begin{table*}[ht]
    \centering
    \fontsize{10}{10}\selectfont
\begin{tabular}{lp{10cm}}
\toprule
subreddit   &  closest subreddits \\
\midrule
        \subreddit{vegan} & \subreddit{intersex}, \subreddit{transgenderUK}, \subreddit{TransSpace}, \subreddit{Transmedical}, \subreddit{ABCDesis}, \subreddit{LGBTindia}, \subreddit{DeepThoughts}, \subreddit{nonduality}, \subreddit{exvegans}, \subreddit{BlockedAndReported}\\
        
        \subreddit{carnivore} & \subreddit{carnivorediet}, \subreddit{Psoriasis}, \subreddit{PsoriaticArthritis}, \subreddit{rheumatoid}, \subreddit{acne}, \subreddit{Hashimotos}, \subreddit{Testosterone}, \subreddit{PlasticSurgery}, \subreddit{lupus}, \subreddit{kratom}\\
        
        \subreddit{communism} & \subreddit{InformedTankie}, \subreddit{CommunismWorldwide}, \subreddit{CPUSA}, \subreddit{socialism}, \subreddit{StupidpolEurope}, \subreddit{LatinAmerica}, \subreddit{sendinthetanks}, \subreddit{ROI}, \subreddit{Africa}, \subreddit{myanmar}\\
         
        \subreddit{Capitalism} & \subreddit{georgism}, \subreddit{Unions}, \subreddit{SandersForPresident}, \subreddit{labor}, \subreddit{NewDealAmerica}, \subreddit{BernieSanders}, \subreddit{WorkersStrikeBack}, \subreddit{ndp}, \subreddit{theydidthemath}, \subreddit{MayDayStrike}\\
        
        \subreddit{monarchism} & \subreddit{leftistvexillology}, \subreddit{UsefulCharts}, \subreddit{RoughRomanMemes}, \subreddit{IndiaPlace}, \subreddit{pureasoiaf}, \subreddit{MedievalHistory}, \subreddit{shittyhalolore}, \subreddit{MemriTVmemes}, \subreddit{ancientrome}, \subreddit{darkwingsdankmemes}\\
        
        \subreddit{AbolishTheMonarchy} & \subreddit{Sham\_Sharma\_Show}, \subreddit{forwardsfromgrandma}, \subreddit{WordAvalanches}, \subreddit{Malaphors}, \subreddit{Lal\_Salaam}, \subreddit{Metal}, \subreddit{BanVideoGames}, \subreddit{QanonKaren}, \subreddit{PoliticalHumor}, \subreddit{insanepeoplefacebook}\\
        
        \subreddit{Millennials} & \subreddit{GenZ}, \subreddit{PlusSizedAndPregnant}, \subreddit{ThailandTourism}, \subreddit{indiasocial}, \subreddit{Liverpool}, \subreddit{TallGirls}, \subreddit{Psychedelics}, \subreddit{Chandigarh}, \subreddit{Cardiff}, \subreddit{ChronicIllness}\\
        
        \subreddit{GenZ} & \subreddit{Millennials}, \subreddit{indiasocial}, \subreddit{TallGirls}, \subreddit{Kibbe}, \subreddit{Zillennials}, \subreddit{aggretsuko}, \subreddit{precure}, \subreddit{nerdfighters},  \subreddit{KoeNoKatachi}, \subreddit{comiccon}\\
        
        \subreddit{GenX} & \subreddit{lesbianfashionadvice}, \subreddit{CosplayNation}, \subreddit{Xennials}, \subreddit{bigboobproblems}, \subreddit{feminineboys}, \subreddit{PlusSize}, \subreddit{distantsocializing}, \subreddit{CasualPH}, \subreddit{bald}, \subreddit{TallGirls}\\
        
        \subreddit{Feminism} & \subreddit{antiwoke}, \subreddit{MensLib}, \subreddit{ControversialOpinions}, \subreddit{EnoughIDWspam}, \subreddit{prochoice}, \subreddit{prolife}, \subreddit{fourthwavewomen}, \subreddit{TransSpace}, \subreddit{TrueUnpopularOpinion}, \subreddit{AntiVegan}\\
        
        \subreddit{MensRights} & \subreddit{antifeminists}, \subreddit{AntiHateCommunities}, \subreddit{AsABlackMan}, \subreddit{LeftWingMaleAdvocates}, \subreddit{aznidentity}, \subreddit{EnoughPCMSpam}, \subreddit{FragileWhiteRedditor}, \subreddit{BlatantMisogyny}, \subreddit{TheLeftCantMeme}, \subreddit{ForwardsFromKlandma}\\
        
        \subreddit{atheism} & \subreddit{exmuslim}, \subreddit{antitheistcheesecake}, \subreddit{extomatoes}, \subreddit{Antitheism}, \subreddit{atheismindia}, \subreddit{mormon}, \subreddit{progressive\_islam}, \subreddit{Judaism}, \subreddit{RadicalChristianity}, \subreddit{religiousfruitcake}\\
        
        \subreddit{spirituality} & \subreddit{SpiritualAwakening}, \subreddit{awakened}, \subreddit{pureretention}, \subreddit{Stoicism}, \subreddit{Shamanism}, \subreddit{starseeds}, \subreddit{Mediums}, \subreddit{Soulnexus}, \subreddit{Semenretention}, \subreddit{SASSWitches}\\
        
        \subreddit{religion} & \subreddit{AskAChristian}, \subreddit{ChristianUniversalism}, \subreddit{RadicalChristianity}, \subreddit{bahai}, \subreddit{hinduism}, \subreddit{shia}, \subreddit{Bible}, \subreddit{progressive\_islam}, \subreddit{extomatoes}, \subreddit{mormon}\\

       \bottomrule
    \end{tabular}
    \caption{For each subreddit in the controversial topics analysis, the 10 most similar subreddits in terms of values.}
    \label{tab:app:controversial}
\end{table*}

\section{Reproducibility}
\label{app:reproducibility}

\subsection{Training the Schwartz Values Extractor}

\label{app:extractor_training_details}
We trained the relevance model and the stance model for ten epochs with early stopping, using a learning rate of $5 \cdot 1e^{-5}$, batch size of 32, Adamw optimiser with default parameters and linear learning rate scheduler. We trained the models on a single TitanRTX GPU for about 5 hours for the relevance model, and 2 hours for the stance model.

\end{document}